\documentclass[reprint,amssymb,amsmath,aip,cha]{revtex4-1}
\usepackage{graphicx}
 \usepackage{textcomp}
\usepackage[LGRgreek]{mathastext}
\usepackage[caption = false]{subfig}
\usepackage{color}
\usepackage[export]{adjustbox}
\usepackage{epsfig}
\usepackage{gensymb}
\usepackage{ifpdf}
\usepackage{bm}
\usepackage[colorlinks=true,linkcolor=blue]{hyperref}
\expandafter\ifx\csname package@font\endcsname\relax\else
 \expandafter\expandafter
 \expandafter\usepackage
 \expandafter\expandafter
 \expandafter{\csname package@font\endcsname}%
\fi
\usepackage{cleveref}
\begin{document}
\title{Activated carbon-assisted anionic dye decomposition
using a dual-liquid post-dielectric barrier discharge
plasma reactor}
\author{Krishna V S}
\affiliation{Department of Physics, Mar Thoma College for Women, Perumbavoor, Ernakulam, Kerala, India}
\author{Surya}
\affiliation{Department of Physics and Astrophysics, University of Delhi, Delhi-110007, India}
\author{Mangilal Choudhary}
\email{mchoudhary@physics.du.ac.in}
\affiliation{Department of Physics and Astrophysics, University of Delhi, Delhi-110007, India}
\begin{abstract}
Non-thermal Plasma (NTP) technology, including the Dielectric Barrier Discharge (DBD) reactor, has emerged as an effective advanced oxidation process (AOP) for treating persistent organic pollutants in industrial wastewater. Based on this objective, a dual-liquid dielectric-barrier-discharge plasma reactor was built and used to degrade a ternary anionic dye mixture comprising Reactive Red 120, Congo Red, and Methyl Orange in its post-discharge configuration. The adjustable length of the outer liquid electrode (grounded) provides flexibility in controlling the discharge length and cooling, thereby achieving higher degradation efficiency and greater reactor stability. The conducting-liquid high-voltage electrode (a dielectric tube filled with a conductive solution) helps mitigate metal corrosion and oxidation caused by reactive species in the discharge zone. A detailed study of the degradation of ternary anionic dye mixtures (simulated dye wastewater samples) in the presence of activated carbon (AC) was performed to explore the potential applications of post-discharge treatment. The activated carbon (granules) enhances the degradation rate of ternary anionic dye mixtures in basic solutions rather than in acidic or neutral solutions. A comparative study of the degradation reaction kinetics of the ternary anionic dye mixtures in acidic and basic solutions was conducted after analyzing UV-Vis absorption spectra. The role of other parameters, such as solution pH, the mass and size of AC particles, dye concentrations, and AC reuse, was investigated in relation to degradation efficiency. The degradation mechanism was elucidated by the formation of strong oxidizer (hydroxyl radicals) on the surface of activated carbon during post-discharge treatment. However, in acidic or neutral anionic dye solutions, the degradation mechanism was dominated by ozone.
\\
\vskip 2mm
\end{abstract}
\maketitle
\maketitle
\date{10.06.2026}
\section{Introduction}
\begin{figure*}[t]
    \centering
    \includegraphics[width=0.95\textwidth]{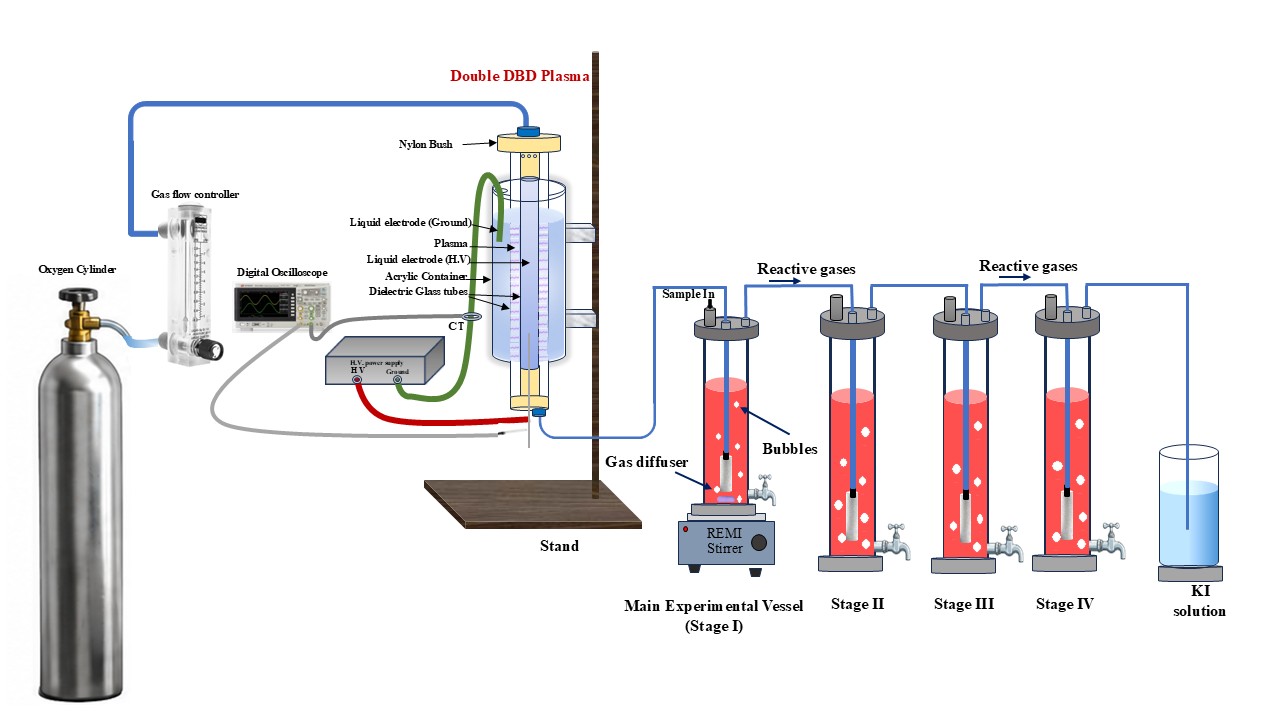}
    \caption{{Diagram of a co-axial cylindrical dual-liquid post-dielectric barrier discharge plasma reactor}}
    \label{fig:Fig1}
\end{figure*}
Water pollution from industrial effluents constitutes one of the most pressing environmental challenges of the twenty-first century. The textile industry, among the world’s largest industrial consumers of freshwater, outflows approximately 200,000 tonnes of synthetic dyes into water bodies annually, representing 10–15\% of total dye production lost during manufacturing, dyeing, and finishing processes \cite{waterpollution,waterpollution2}.  The practical reality of textile dyeing operations is that the effluents contain multiple co-existing dye classes. A single dye class simultaneously comprises reactive dyes for cellulose fibres, direct dyes for cotton and viscose, and acid dyes for wool and nylon, resulting in effluents with complex ternary or higher-order dye mixtures that are more complex than any individual component. If effluents from the textile industry are not treated before discharge into water bodies, high concentrations of dyes can harm aquatic animals and plants\cite{effectdyeolivingorganism_19,dye_effect_health_2003}. 
\\
The challenges in treating dye effluent (containing anionic dyes) with physio-chemical and biochemical processes motivate research around the globe to explore advanced dye degradation methods such as advanced oxidation methods\cite{advancedoxidation1, advancedoxidation2,advancedoxidation4}, photo-catalyst techniques\cite{photodegradtaionreview2024,uv_photo_degradation_2008}, and biological treatment\cite{biologicalmethod1,biologicalmethod2,biologicaldye_muthu2022_book}.  
In recent decades, the potential applications of Non-Thermal Plasma (NTP) technology, which is considered an advanced oxidation process, along with other existing methods for decomposing synthetic dyes, have been explored in many laboratories\cite{needle-plate_dyedegradation_2016,nspulsedcorona_dydegradation_2019,multineedleunderwaterbubble_MB2024,plasmajet_dyedegrade_satya_2022,plasmawaterinteractionreview1, hybrid_corona_dbd_2021}. The dielectric barrier discharge (DBD) plasma reactor (NTP configuration) can be used in direct mode (plasma-water interaction) \cite{DBDconfiguration_diffrent_degradtaion_review15,dbd_waterdroping_dyedegrade_2013,surfacebarrier_dbd_dyedegrade_2022} and indirect mode (post-discharge treatment)\cite{postdischarge_gasphasedbd_dyedegradtaion_2015,postdbd_degradtaion_CV_2019,mangi_post-discharge_2026}. The non-thermal atmospheric pressure plasmas contain various reactive oxygen and nitrogen species (RONS), including hydroxyl radicals ($\cdot$OH), ozone ($\text{O}_3$), hydrogen peroxide ($\text{H}_2\text{O}_2$), superoxide radicals, nitrite ($\text{NO}_2^-$), and nitrate ($\text{NO}_3^-$), along with energetic electrons and ultraviolet photons \cite{reactivespecies_discharges_review_2018}. The strong oxidizers ($\text{O}_3$ and $\cdot$OH) and energetic electrons decompose the synthetic dye molecules during the plasma-wastewater interaction\cite{plasmawaterinteractionreview1}. However, long-lived oxidizers ($\text{O}_3$ and $\text{H}_2\text{O}_2$) play a dominant role in degrading the dye molecules in post-discharge configurations\cite{mangi_post-discharge_2026}. \\
As stated in the previous work \cite{mangi_post-discharge_2026}, the degradation efficiency is very low in direct plasma-wastewater treatment due to the smaller plasma-water interaction area. Therefore, the investigation into integrating post-discharge treatments (advanced oxidation) with suitable catalysts and UV radiation to decompose toxic organic compounds (dyes and their toxic fragments) in water solutions (textile industrial wastewater) is essential for healthy environments.\\
One of the synthetic dye groups, anionic dyes (containing azo groups and sulfonate groups), is widely used in the textile and printing industries, and they are considered a significant class of water pollutants because of their high chemical stability and persistence in aqueous environments, which allows them to remain in water systems for longer periods and pose toxicity risks to aquatic organisms. Moreover, anionic dyes are difficult to remove from industrial wastewater due to their higher water solubility, electrostatic repulsion with traditional adsorbents, and resistance to biological degradation. Under anaerobic conditions, cleavage of the azo bond may produce toxic aromatic amines that are often more hazardous than the parent dye molecules themselves. Since anionic dyes possess a wide range of structural diversity, they create major differences in degradation behaviour, making the treatment of real industrial effluents highly challenging using conventional methods\cite{chemicalmethod2, physical_methods_review, biologicaldye_muthu2022_book}.\\\\
As it has been highlighted that dielectric barrier discharges have great potential to create short-lived strong oxidizers ($\cdot$OH, super-oxides, etc.) as well as long-lived strong oxidizers ($O_3$ and $H_2O_2$) at atmospheric pressure. Therefore, the interaction of long-lived oxidizers generated during oxygen breakdown with anionic dyes has been considered an effective means of their decomposition due to their high oxidative ability, use without additional chemicals, and environmental acceptability. However, the limited gas-liquid mass transfer of oxidizers in the post-discharge configuration (Fig.~\ref{fig:Fig1}) reduces the degradation efficiency of anionic dyes. Therefore, the long-lived oxidizers emanating from the dielectric barrier discharge zone (discharge region) are dispersed into the simulated anionic dye mixture solution (a model textile wastewater sample) via gas diffusers (bubblers) to enhance the gas-liquid mass transfer rate\cite{postdbd_degradtaion_CV_2019,mangi_post-discharge_2026}. Another way to utilize the emanating oxidizers in the gas phase is to use a sequential, multi-stage treatment process (see Fig.\ref{fig:Fig1}) rather than a single stage to enhance the overall degradation efficiency of the plasma reactor. A detailed study of the advantages of multi-stage configurations in post-DBD has been conducted separately and is under review. It has also been demonstrated in previous studies that the nature (acidic or basic) of dye solutions or wastewater strongly affects the degradation rate during the oxidation processes. This suggests that the degradation efficiency of anionic dyes in solutions at different pH values may increase or decrease during post-discharge treatment. The degradation of an anionic dye and its mixture at different pH levels, in the presence of a long-lived oxidizer ($O_3$) and activated carbon (granules) as a catalyst, was the subject of this study.\\\\
In a normal DBD plasma reactor \cite{postdbd_degradtaion_CV_2019}, used to degrade dyes through advanced oxidation processes, metal electrodes (stainless steel, aluminium, or copper) are used as a high voltage electrode and a grounded (cathode) electrode. Air or oxygen can be used as a working gas in the DBD reactor to generate strong oxidizers to decompose organic dyes or organic pollutants. In our previous study, we highlighted the novelty of replacing the outer wire mesh or a metal sheet (grounded) with a conducting liquid electrode (grounded)\cite{mangi_post-discharge_2026}. There are still challenges in using aluminium, stainless steel, or brass as a high-voltage electrode in a coaxial cylindrical DBD reactor. The major problem is the formation of oxide layers on the high-voltage electrode due to interactions among various strong oxidizers in the discharge zone, which reduces the stability and performance of the DBD reactor. \\
To overcome these challenges, a low-cost new dual-liquid post-double dielectric barrier discharge plasma reactor was designed and developed in the plasma physics laboratory. In this newly developed double dielectric barrier discharge (DDBD), plasma is generated between two dielectric layers (tube walls), which are in contact with the conducting liquids (potassium chloride, KCl), as depicted in Fig.~\ref{fig:Fig1}. The discharge (plasma) length can be adjusted by changing the height of the outer conducting liquid (grounded electrode) to optimize the reactor for higher energy yield. The role of the outer conducting liquid is to act as an outer electrode as well as a coolant to stabilize and ensure safe operation of the DDBD reactor for a longer duration.\\\\
The newly built dual-liquid DDBD reactor was utilized to study the decolourization (degradation) of anionic ternary mixtures (RR 120, MO and CR) as model dye wastewater samples in the absence and the presence of activated carbon (catalyst). First, the discharge length was optimized to achieve maximum dye decomposition at a given oxygen flow rate and magnetic stirrer speed. The degradation efficiency and reaction kinetics of anionic ternary mixtures at different pH levels (acidic and basic solutions) were investigated in the post-DDBD configuration, both in the absence and presence of activated carbon (granules). The concentrations of long-lived reactive species in the gas emanating from the dual-liquid DDBD reactor (post-discharge gases) were estimated using analytical techniques to explore the degradation mechanism (catalytic ozone decomposition) in the presence of activated carbon particles. The impact of operating factors, including the concentration of dye solution, gas flow rates, size and mass load of activated carbon particles, usage cycles of activated particles and the role of oxygen on the degradation (decolourization) process was also examined. \\\\
The remaining paper is organized as follows: The dual-liquid post-DDBD plasma reactor and the instruments used in this anionic dye degradation are discussed in Sec.~\ref{sec:sec2}. Experimental procedure, optimization of length, and analytical techniques are discussed in Sec.~\ref{sec:sec3}. The experimental results on the anionic ternary mixtures (acidic and basic), with and without activated carbon, are presented in Sec.~\ref{sec:sec4}. The observed results are discussed in Sec.~\ref{sec:sec5}. Sec .~\ref{sec:sec6} summarizes the reported work and directs possible future directions. 
\section{Experimental setup and instruments} \label{sec:sec2}
\begin{figure*}[t]
    \centering
    \begin{minipage}[t]{0.33\textwidth}
        \centering
        \includegraphics[width=\linewidth]{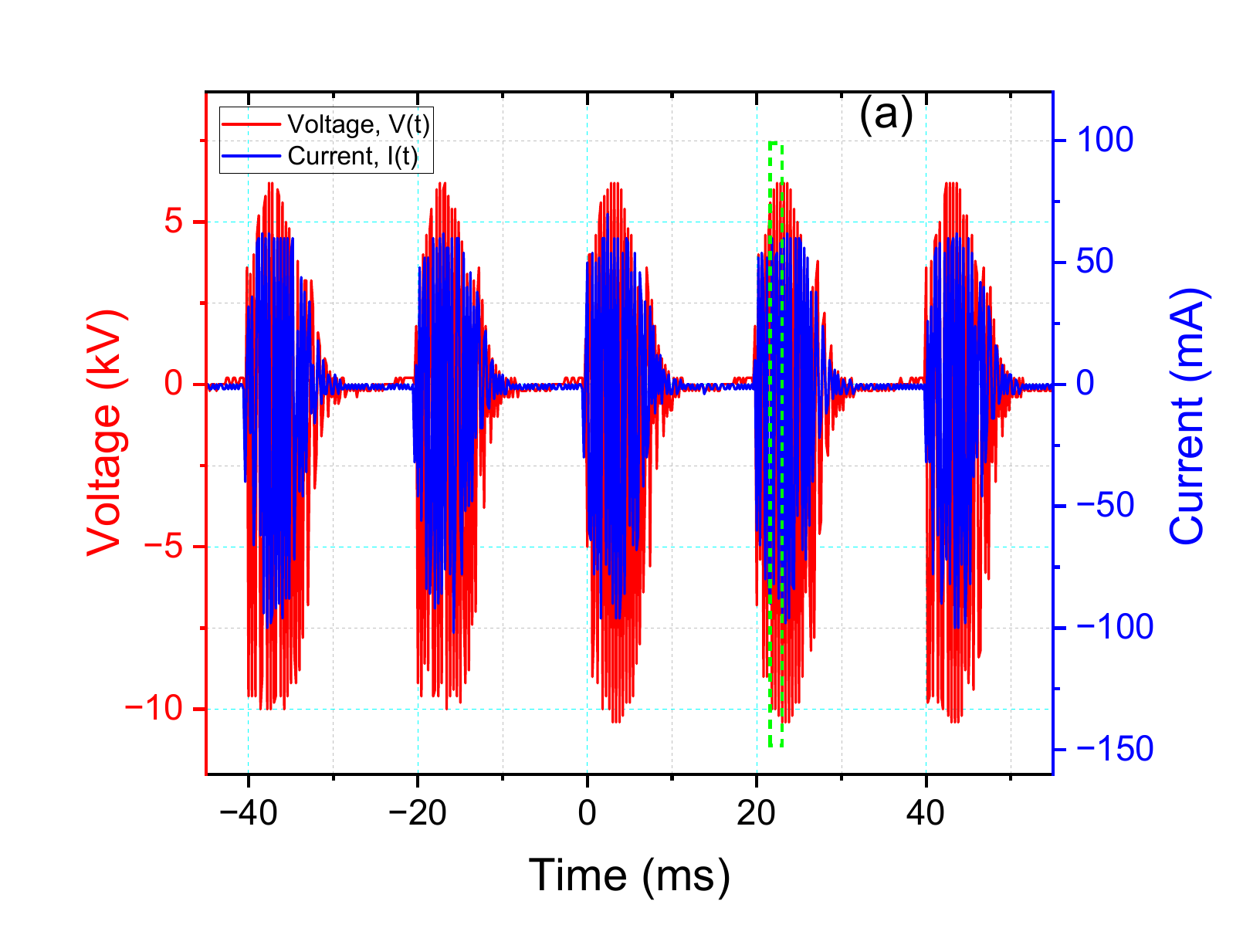}
    \end{minipage}%
    \hfill
    \begin{minipage}[t]{0.33\textwidth}
        \centering
        \includegraphics[width=\linewidth]{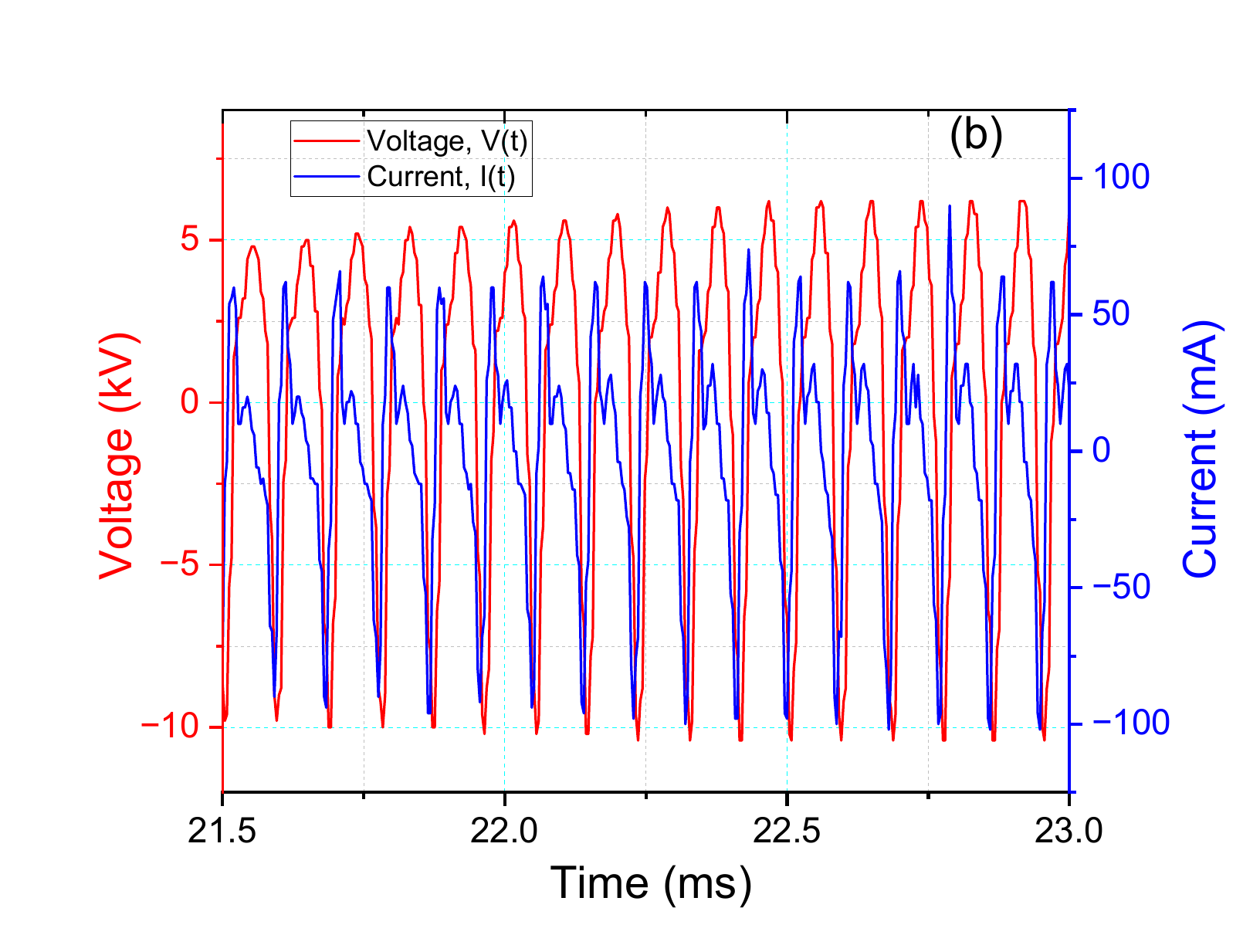}
    \end{minipage}%
    \hfill
    \begin{minipage}[t]{0.33\textwidth}
        \centering
        \includegraphics[width=\linewidth]{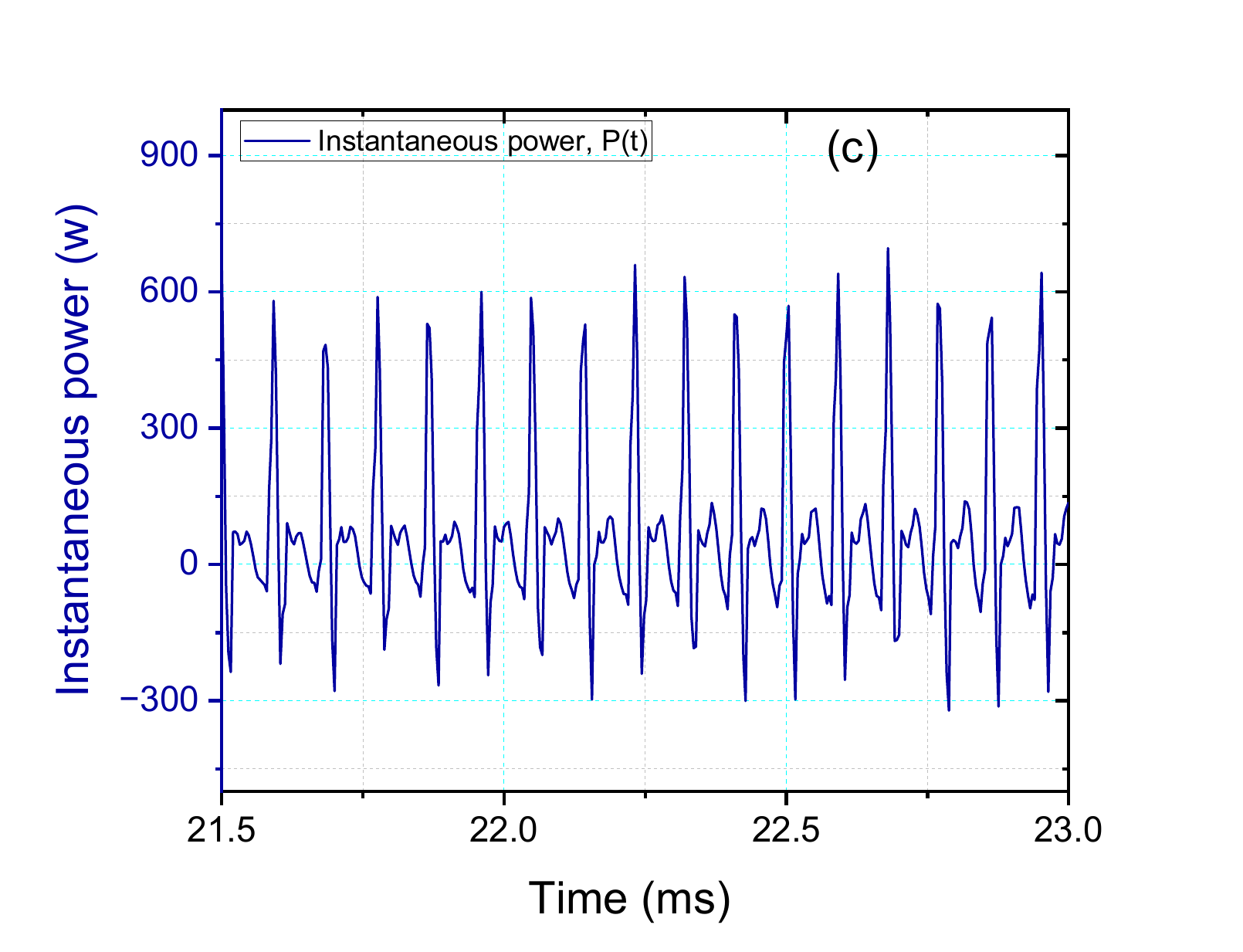}
    \end{minipage}

    \caption{{(a) I--V characteristics, (b) I--V characteristics (zoomed in view of marked region in Fig.\ref{fig:Fig2}(a)), and (c) instantaneous dissipative power $P(t)$.}}
    \label{fig:Fig2}
\end{figure*}
A schematic diagram of an in-house-built, low-cost coaxial cylindrical dual-liquid double-dielectric barrier discharge (DDBD) reactor is shown in Fig.\ref{fig:Fig1}. In this coaxial DDBD configuration, a conducting liquid (Potassium chloride, KCl) filled dielectric tube (borosilicate glass) with an inner diameter (I.D.) of 12 mm and an outer diameter (O.D.) of 16 mm was fixed into a 19 mm inner diameter and 22 mm outer diameter borosilicate tube through specifically designed nylon feed-through bushes. A stainless-steel high-voltage electrode (2 mm diameter) was kept inside the KCl-filled dielectric tube and connected to an in-house-built high-voltage AC power supply (26 kV, 40 mA). The discharge region was formed in the annular gap of 3 mm between the inner dielectric tube (liquid electrode) and the outer borosilicate glass tube, where oxygen ($\sim$ 99\% purity) was introduced through a gas connector fitted on the nylon bush (see Fig.\ref{fig:Fig1}). This assembly of concentric dielectric tubes was fixed into an acrylic tube filled with KCl (a conducting liquid), with inner and outer diameters of 55 mm and 65 mm, respectively. The outer surface of a large-diameter borosilicate glass tube (O.D. 22 mm) was in contact with KCl solution, serving as a liquid electrode (grounded electrode) and a coolant. The gas breakdown occurs in the air gap between both dielectric surfaces (double dielectric) and liquid electrodes (dual-liquid electrode) once the potential difference crosses 
a threshold value (breakdown voltage). The level of the KCl solution in the acrylic tube could be adjusted externally to vary the effective grounded-electrode length (critical length) and, thereby, the plasma discharge length. This arrangement provided flexibility in controlling plasma generation while keeping the discharge gap fixed. The transient discharge characteristics at a fixed input power and a gas pressure were monitored using a high-voltage 1000X probe and a current transformer.\\\\
The reactive gases emanating from the discharge zone, along with oxygen, were allowed to disperse into the anionic dye solution, which was filled into an acrylic tube (I.D. 80 mm, length 200 mm, volume 1000 ml). During gas-liquid interaction, a gas diffuser (bubbler) and a magnetic stirrer (REMI 2L) were used to enhance the mass transfer rate for the degradation of the dye (an organic pollutant).  The non-dissolved gases emanating (reactive oxidizers and oxygen) at the first stage (main experimental vessel) of treatment were transferred to the remaining stages (stage-II, stage-III, and stage-IV), which were interconnected in series via a polyurethane (PU) tube with an inner diameter of 4 mm and a length of 30 mm (see Fig.~\ref{fig:Fig1}). To prevent backflow of fluid, unidirectional gas valves were installed between connecting PU pipes at each stage. The purpose of using multiple stages in this post-discharge configuration was to utilize the non-dissolved oxidizers ($O_3$ and $H_2O_2$) emanating at each stage of treatment for a simultaneous laboratory study of different wastewaters (synthetic) treatment under varying conditions, such as the role of pH, different catalysts, UV radiation, etc. 
After untilization of oxidizers (reactive gases) at four different stages, the exhaust pipe of the last vessel (stage IV) was connected to a 500 ml potassium iodide ($KI$) solution to neutralize the ozone, which is not good for the environment. All the experimental dye solution vessels were made of acrylic material and were vacuum-tight. The dimensions of the other acrylic containers (I.D. 55 mm, length 500 mm) differed from those of the main experimental vessel. We had a provision to inject the activated carbon particles (catalyst) and collect synthetic wastewater samples during treatment in each vessel. For this present study, we conducted a series of experiments in the main experimental vessel, with or without activated carbon. However, other stages were used in simultaneous studies on different dye solutions under different conditions, which are not a part of these findings.\\  
The scientific-grade anionic dyes, MO (SRL make), CR (SRL make), and RR 120 (SRL make) were purchased from an authorized vendor.  The pH, electrical conductivity (EC), and total dissolved solids (TDS) of synthetic dye wastewater were measured using the corresponding meters. A UV-Vis spectrometer (PerkinElmer Lambda 950) was used to analyze the degradation kinetics of individual as well as ternary mixtures of anionic dyes in the presence or absence of activated carbon. The concentration of the dominant oxidizer ($O_3$) in post-discharge reactive gases was measured using the iodometric titration method. Other reactive species in emanating gas were measured using the analytical test kits.  
\section{Experimental Procedure and analytical techniques} \label{sec:sec3}
\begin{figure}
   \centering
   \includegraphics[width=0.95\linewidth]{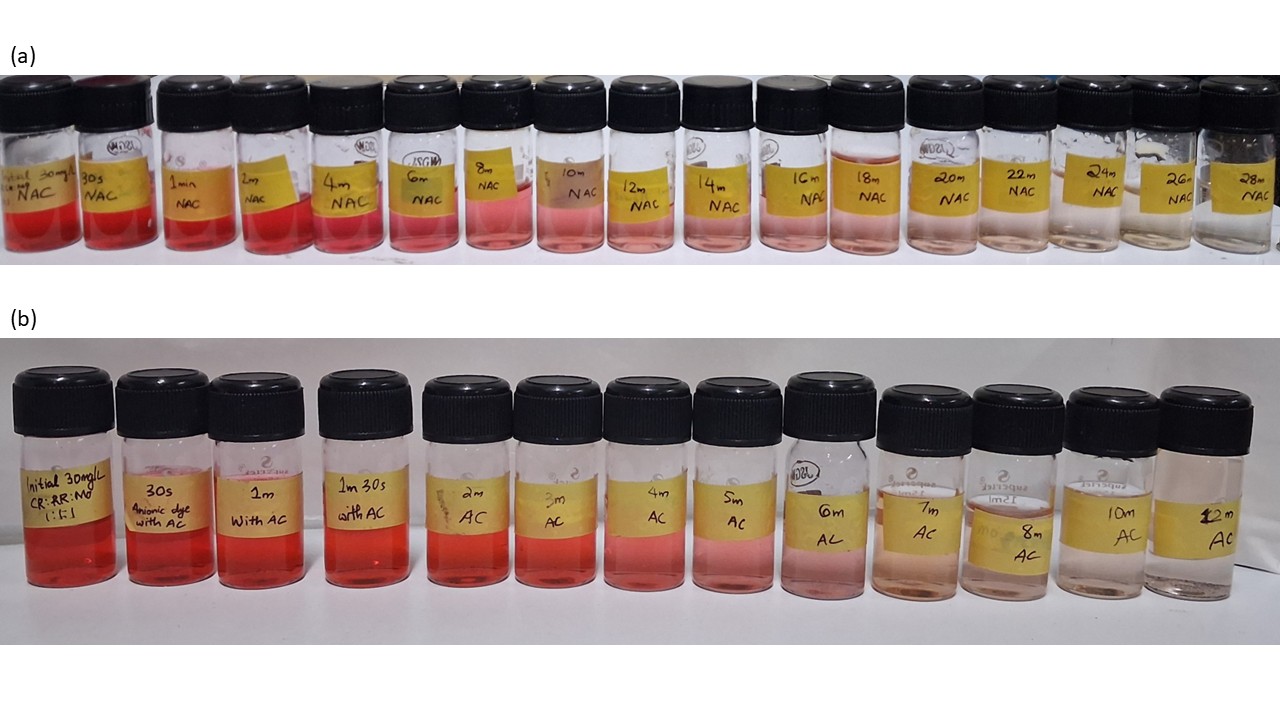},
    \caption{\textit{}{Decolourization of symmetric ternary anionic dye mixture (30 mg/L) in alkaline conditions (a) without activated carbon and (b) with activated carbon (2 g) during post-discharge treatment.}}
    \label{fig:Fig3}
\end{figure}

\begin{figure}
   \centering
   \includegraphics[width=0.93\linewidth]{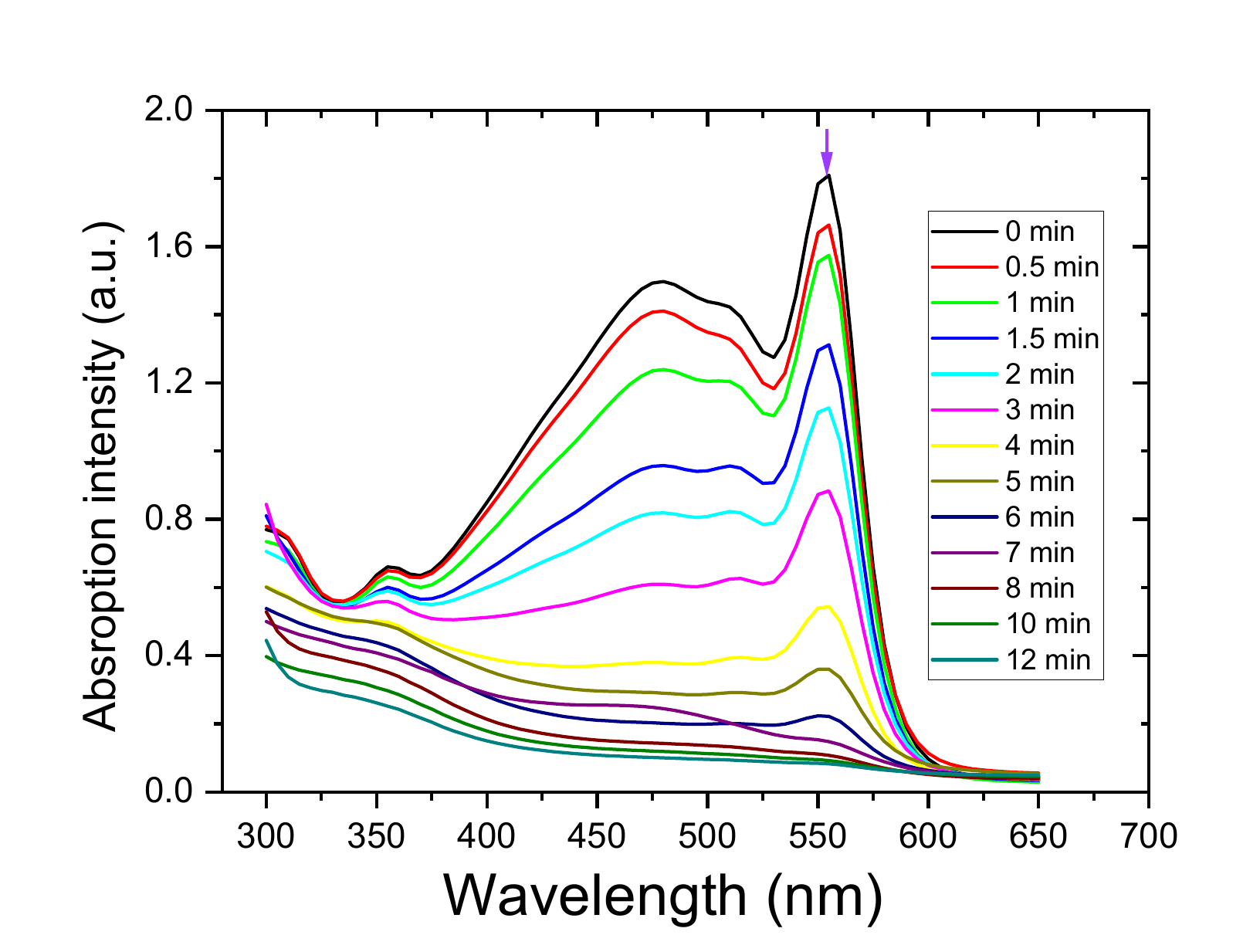},
    \caption{\textit{}{UV-Vis absorption spectrum of symmetric ternary anionic dye mixture (30 mg/L) [MO (10mg/L), CR (10mg/L) and RR120 (10mg/L)] in alkaline conditions with activated carbon (granules) at different post-discharge treatment times. The volume of the ternary anionic dye mixture solution was 800 ml in the main experimental chamber, the gas flow rate was 300 cc/min, and the stirrer speed was 400 RPM. An arrow indicates the absorption peak at the given time.}}
    \label{fig:Fig4}
\end{figure}

\begin{figure}
   \centering
   \includegraphics[width=0.95\linewidth]{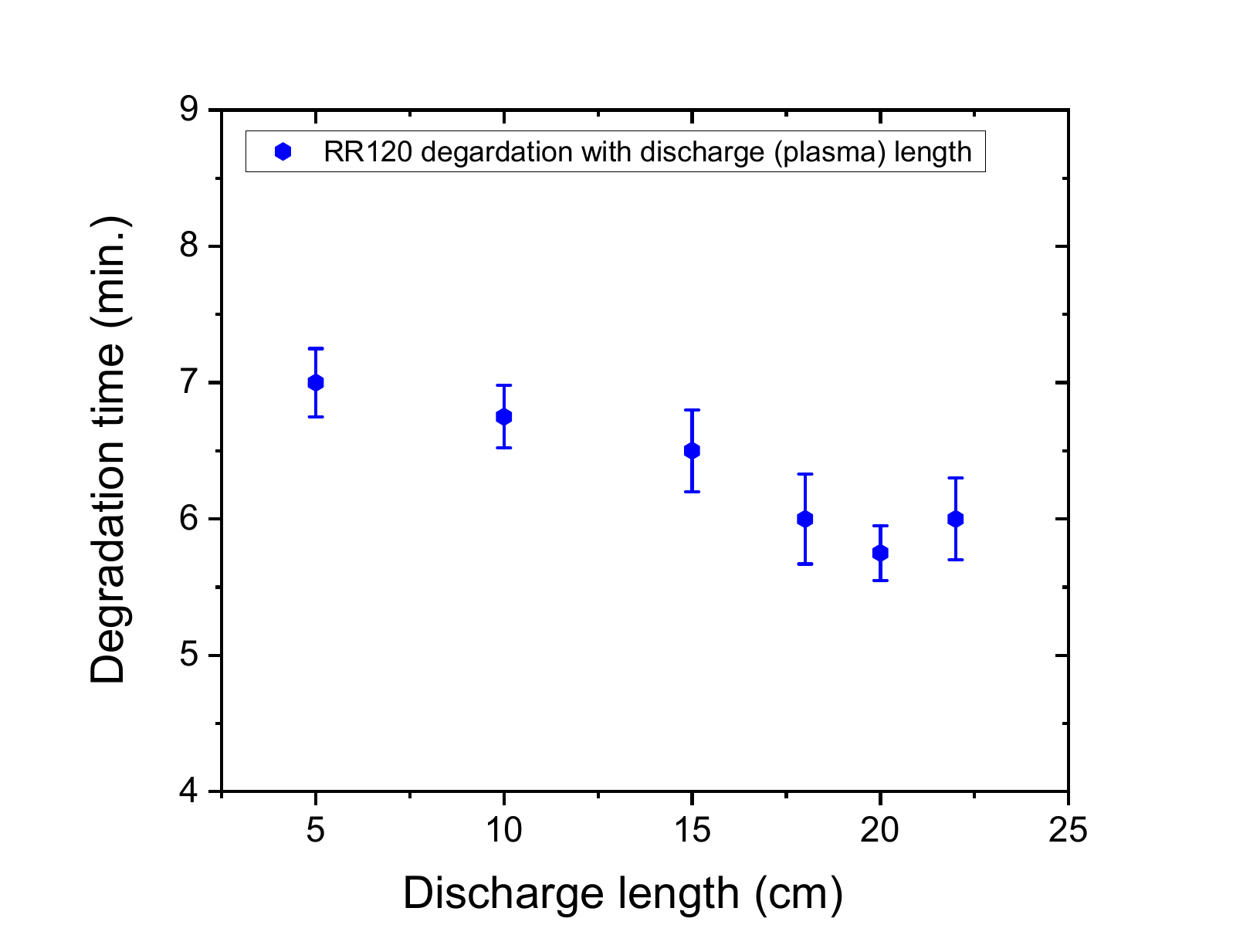},
    \caption{\textit{}{Variation of degradation ($>$ 95 \%) time for RR 120 (30 mg/L) at different discharge lengths. The volume of dye solution, gas flow rate, and speed of the magnetic stirrer were 800 ml, 400 cc/min and 400 RPM, respectively.}}
    \label{fig:Fig5}
\end{figure}

The gas breakdown occurs in the annular region between the dielectric surfaces when a high-voltage signal (50 Hz, 50\% duty cycle) is applied to the liquid electrodes. This indicates plasma formation only during the positive cycle of the 50 Hz amplified voltage signal at the power supply's output terminal. There was a provision on the power supply to control the ON and OFF times of the applied voltage to the electrodes. In the present study, we set the ON and OFF times to 50\% (3 seconds ON, 3 seconds OFF) of the total exposure time to avoid damage to the electrical circuits and thermal destabilization of the reactor. The voltage signal ($V(t)$) at the inner liquid electrode (anode) and the corresponding transient discharge current ($I(t)$) flowing in the circuit during gas breakdown are depicted in Fig.\ref{fig:Fig2}(a). The sharp spikes in voltage at the electrode and the corresponding discharge current are due to the formation of micro-discharges (short-lived current filaments) between dielectric surfaces. In the 10 ms discharge time span (positive cycle of the applied voltage signal), the recorded momentary voltage ($V(t)$) and discharge current ($I(t)$) signals are shown in Fig.\ref{fig:Fig2}(b). The momentary currents  (Fig.\ref{fig:Fig2}(b)) signal has spikes that correspond to displacement current along with conduction current. Therefore, the transient current signals were filtered before measuring the instantaneous reactive power. \\
The instantaneous dissipative reactive power ($P(t)$ = $V(t) \times I(t)$) based on momentary values of voltage ($V(t)$) and filtered discharge current (conduction current) is plotted in Fig.\ref{fig:Fig2}(c). The average dissipative power to the DDBD load for each input voltage signal at 240 V and 50 Hz (at the power supply input) was estimated using the integration method. 
\begin{equation}
    P_{avg} = \frac{1}{T} \int_0^T V(t) I(t)dt
\end{equation}
where T is the time period of a voltage signal at a high voltage electrode. 
The average power loss in maintaining discharge was calculated by integrating the instantaneous power (Fig.\ref{fig:Fig2}(c)) for different discharge cycles (Fig.\ref{fig:Fig2}(a)) over a given time duration. The value of the average power delivered to the DDBD load was between 20 and 25 W for a single input voltage signal (50 Hz).
\\\\
The symmetric ternary anionic dye mixture (RR 120, MO, and CR) was prepared at varying initial concentrations by diluting the high-concentration stock solutions. The nature (acidic or basic) of the ternary anionic dye mixture solution of desired concentration was adjusted by adding the required volume of sulfuric acid ($H_2SO_4$) and potassium hydroxide ($KOH$) of a given concentration (0.1 M). For using activated carbon (granules) as a catalyst, larger-sized activated carbon grains were crushed into smaller-sized (a few mm) poly-dispersed particles. Later, the required amount of AC (2 g, 4 g, or 6 g, etc.) of a given average size was weighed and used as a catalyst in degradation studies.\\
The experiments were conducted by preparing 800 ml of synthetic dye wastewater (1 MO:1 CR:1 RR 120) of a given nature (acidic, neutral, or basic), with or without activated carbon particles, in the main experimental container. The change in colour of dye solutions with post-discharge treatment (with and without AC) is a qualitative indication of the degradation of dye molecules into smaller organic compounds. The degradation of symmetric ternary anionic dye mixtures (30 mg/L) in alkaline solution (pH $>$ 11), with and without activated carbon, is shown in Fig.\ref{fig:Fig3}. We observe an almost transparent dye solution in both cases once more than 90\% of the dye molecules are decomposed. However, for the quantitative decolourization studies of solutions, UV-Vis absorption spectroscopy of treated samples (5 ml) was conducted. A UV-Vis absorption spectrum of symmetric ternary anionic dyes (30 mg/L) in an alkaline environment in the presence of activated carbon is displayed in Fig.\ref{fig:Fig4}. The maximum absorption peaks of the spectrum, proportional to the concentration of dye molecules, were used to estimate the efficiency and type of degradation reaction kinetics. \\
For evaluating a reactor's energy efficiency, the energy yield (Y) for pollutant degradation is a critical metric. The energy yield (Y) is calculated using the following formula:
\begin{equation}
\text{Y(g/kWh)} = \frac{C_{0}\,(g/L) \times V\,(L) \times R\,(\%)}{P\,(kW) \times t\,(h) \times 100}
\end{equation}
where $C_{0}$ is the initial concentration (g/L) of anionic dye mixtures, V is the volume of solution in litres (L), P is the average dissipative power to DBD load (kW), t is the treatment time in hours, and R is the dye (pollutants) degradation percentage (\%) over treatment time t.\\\\ 
Before conducting the degradation study of anionic dye mixtures at different solution conditions, the critical length of the outer liquid electrode (grounded) was optimized. The degradation time of the RR 120 dye solution (30 mg/L, without AC) in the post-discharge configuration was measured by varying the KCl level (liquid-grounded electrode) from 5 cm to 21 cm while maintaining similar experimental conditions. As depicted in Fig.\ref{fig:Fig5}, we see a dip in treatment time ($t$ ~5.5 minutes) at a discharge length of 20 cm, which was considered a threshold (optimized) discharge length in this dual-liquid DDBD plasma reactor.\\\\
\section{Experimental Results} \label{sec:sec4}
The dual-liquid DDBD reactor in a post-discharge configuration was used to study the decomposition of symmetric anionic dye (CR, MO, and RR 120) mixtures under varying experimental conditions, including pH, initial concentration, mass loading of activated carbon, and a reusable cycle of AC.
\subsection{Acidic anionic dye mixtures (symmetric) with and without AC}
To explore the role of activated carbon (granules) in the synthetic dye wastewater (mixture of anionic dyes) treatment under acidic conditions (pH $<$ 7). We conducted a series of experiments on a symmetric ternary dye mixture, with and without activated carbon, in the main experimental vessel. Ternary anionic dye solutions of different concentrations with strong and mild acidic properties were treated, with and without activated carbon (2 g, reusable), in the post-discharge configuration. To explore dye degradation efficiency and reaction kinetics, UV-Vis absorption spectra were analyzed to determine dye concentration at a specific treatment time. \\
The degradation efficiency was estimated as-
\begin{equation}
    \text{Efficiency} (\%) = \frac{C_{0} - C_{t}}{C_{0}} \times 100
\end{equation}
where C$_{0}$, C$_{t}$ are the initial dye concentration and the concentration of the dye solution at a given time t, respectively. 
To understand the rate of dye decomposition with post-discharge treatment, we estimated the order of degradation kinetics as well as the rate constant using the following relations:
\begin{equation}
\text{reaction rate} = - \frac{dC_{t}}{dt} = k_{n} \times [C_{t}]^{n}
\end{equation}
This rate equation can be used to derive the zeroth-, first-, and second-order reaction rate equations after using different values of $n$ (0,1, and 2).
\begin{equation} \label{eq.5}
C_{t} = C_{0} - kt
\end{equation}
\begin{equation}   \label{eq.6}
\ln \left(\frac{C_{t}}{C_{0}}\right) = -kt
\end{equation}
\begin{equation} \label{eq.7}
\frac{1}{C_{t}} - \frac{1}{C_{0}} = kt
\end{equation}
where $k_0$, $k_1$, and $k_2$ are the rate constants for the zero, first, and second order reaction rate equations, respectively. \\\\
\begin{figure*}[t]
    \centering
    \begin{minipage}[t]{0.33\textwidth}
        \centering
        \includegraphics[width=\linewidth]{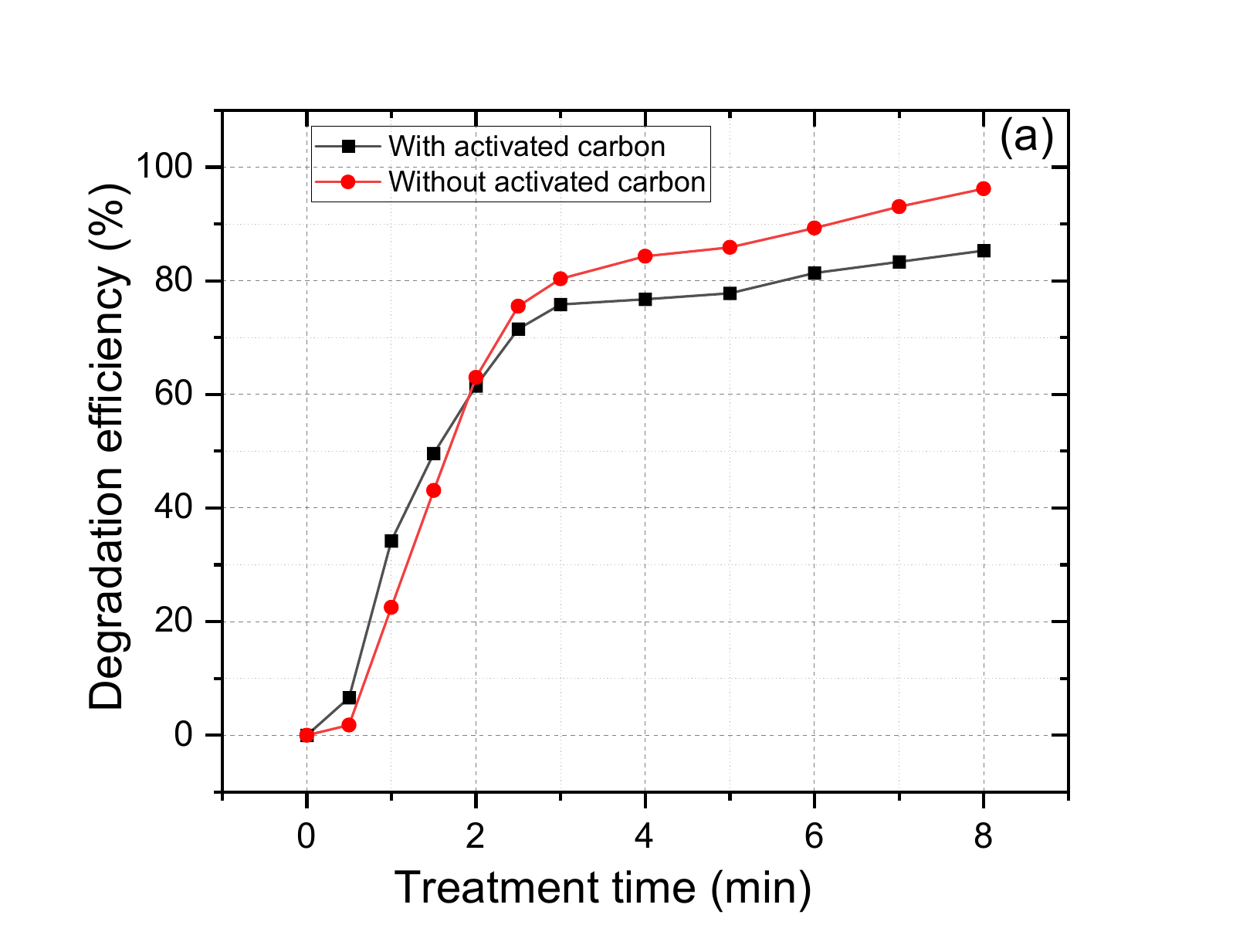}
    \end{minipage}%
    \hfill
    \begin{minipage}[t]{0.33\textwidth}
        \centering
        \includegraphics[width=\linewidth]{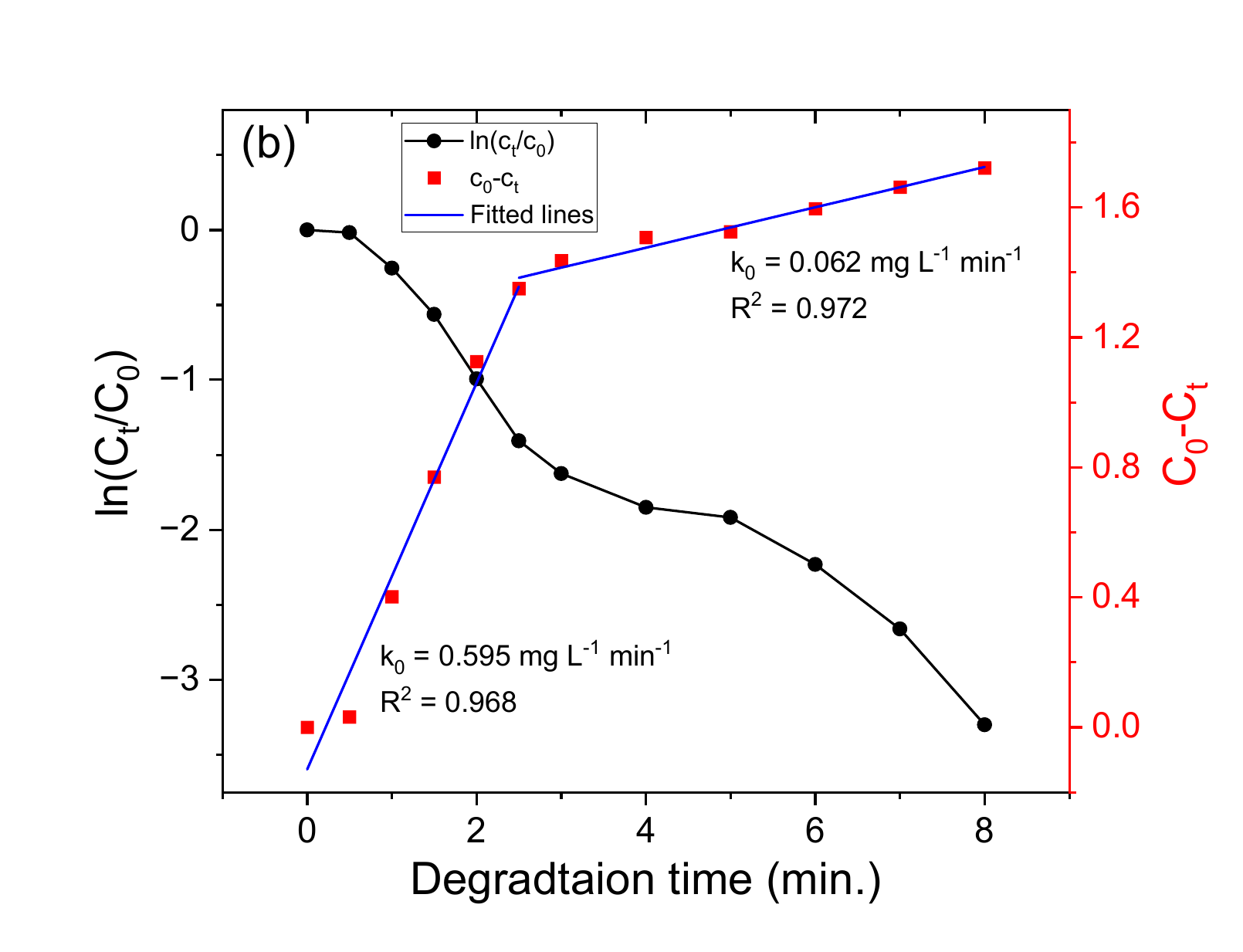}
    \end{minipage}%
    \hfill
    \begin{minipage}[t]{0.33\textwidth}
        \centering
        \includegraphics[width=\linewidth]{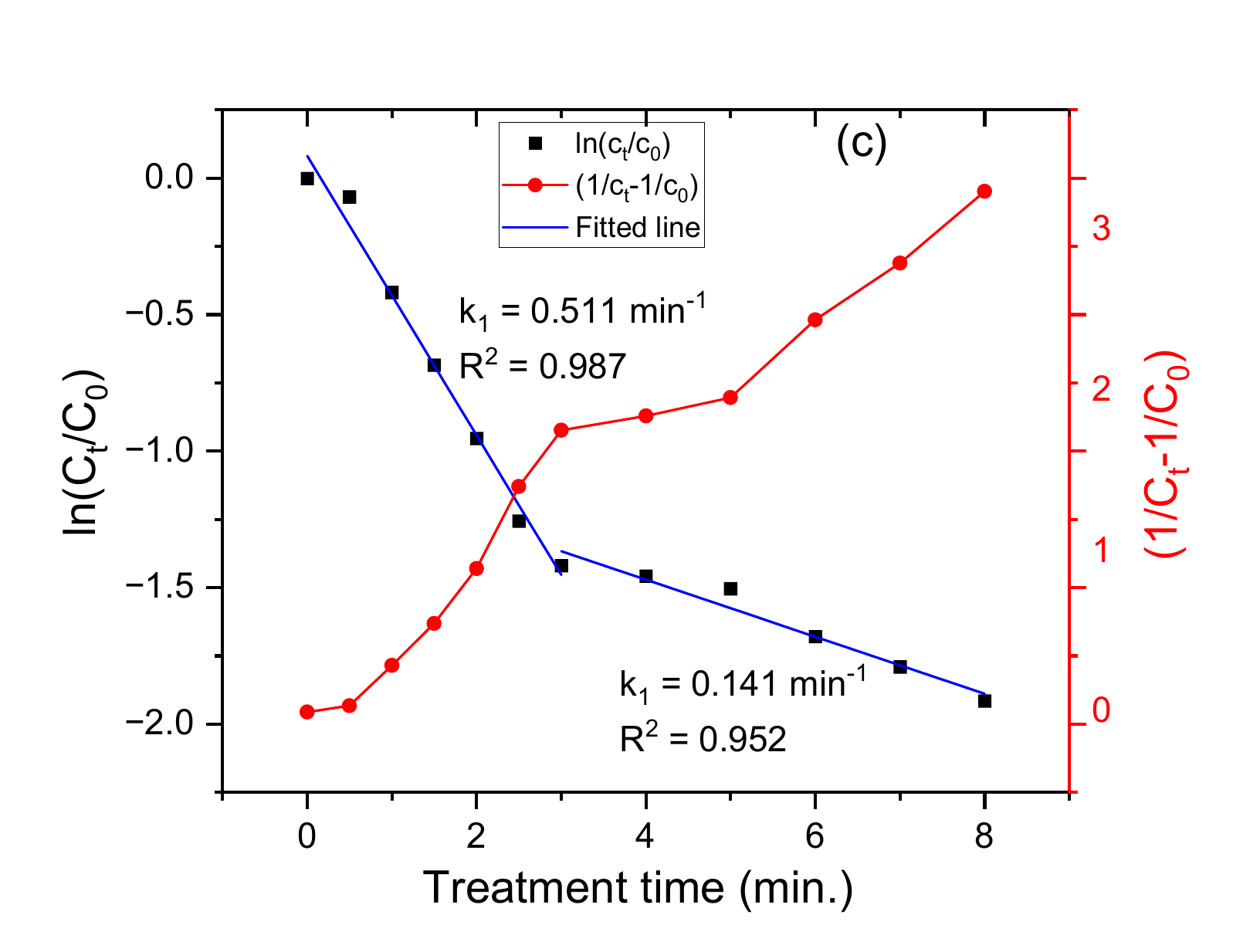}
    \end{minipage}

    \caption{{ (a) Degradation efficiency of the symmetric ternary anionic dye mixture (30 mg/L) under acidic conditions with and without AC. The degradation reaction kinetics of this solution, (b) in the absence of activated carbon, and (c) in the presence of activated carbon. The solution volume, gas flow rate, and magnetic stirrer speed were 800 ml, 300 cc/min, and 400 RPM, respectively.}}
    \label{fig:Fig6}
\end{figure*}
The degradation efficiency of the strongly acidic (pH $\sim$ 1.4) symmetric anionic dyes (MO, CR, and RR 120) solution with a concentration of 30 mg/L, with and without activated carbon, is shown in Fig.\ref{fig:Fig6}(a). The volume of the dye mixture solution was 800 ml, and it was stirred at 400 RPM during the post-discharge treatment with an oxygen flow rate of 300 cc/min. We observed a slight change in degradation efficiency with and without AC up to 80 \%, but beyond that, the decolourization rate was slower in the case of the solution treated in the presence of AC. We observed a similar degradation trend in another set of experiments using anionic dye mixtures under acidic conditions (at different pH values). It should be noted that the UV-Vis absorption spectrum of ternary anionic dyes (with and without AC) is a resultant of individual absorption spectra in the wavelength range of 200 to 800 nm; therefore, the resultant absorption peak ($\sim$ 505 nm) was considered to estimate the degradation efficiency as well as the rate of degradation. However, we assessed the degradation efficiency and reaction kinetics for another absorption peak of the resultant UV-vis spectrum, but did not observe a significant change. \\
The reaction kinetics for each case (with and without AC) were determined by plotting the concentration data against treatment time for zero-, first-, and second-order rate equations (Eq.\ref{eq.5},\ref{eq.6},\ref{eq.7}). Based on the fitted rate equation to the experimental data (dye concentration), the order and the rate constant are confirmed. For a better understanding of the degradation reaction kinetics, concentration data more closely related to the given reaction orders (zero, first, and second) are plotted in each Figure. Fig.\ref{fig:Fig6} (b) shows the degradation reaction kinetics (zero and first order) of a symmetric ternary anionic dye mixture (30 mg/L) without activated carbon. The variation of concentration against treatment time follows the zeroth-order rate reaction with two degradation phases. The rate constant (k$_{0}$ = 0.595 mg$L^{-1}$$min^{-1}$) for the first phase of degradation is nearly 9.6 times greater than the rate constant (k$_{0}$ = 0.062 mg$L^{-1}$$min^{-1}$) for the second phase of degradation. This reduction in the rate constant is expected due to the formation of smaller-sized organic intermediate compounds during the degradation of dye molecules, which reduces the oxidizer's mass transfer rate to dye molecules. \\
For a comparative degradation reaction kinetics study with and without activated carbon, we plotted the dye concentration data corresponding to first and second order rate reactions against treatment time in Fig.\ref{fig:Fig6}(c). We see the best linear fitting of $\ln\left(\frac{C_t}{C_0}\right)$ against the treatment time (Fig.\ref{fig:Fig6}(c)), which indicates that the degradation followed first-order kinetics with two phases (0-2 and 2-8 minutes). The rate constant for the initial phase (k$_{1}$ = 0.511 $min^{-1}$) was relatively much higher than that of the second phase (k$_{2}$ = 0.141 $min^{-1}$). This indicates the initial, faster degradation of the dyes in the presence of AC, which then slows after 2-3 minutes. Different phases of degradation are expected due to a reduction in the availability of oxidizers for the decomposition of dye molecules compared to the intermediate fragments of the degraded dye molecules. There are always possibilities for competing reactions between oxidizers and their scavengers (reaction inhibitors).\\
\subsection{Basic anionic dye mixtures (symmetric) with and without AC}
\begin{figure*}[t]
    \centering

    \begin{minipage}[t]{0.33\textwidth}
        \centering
        \includegraphics[width=\linewidth]{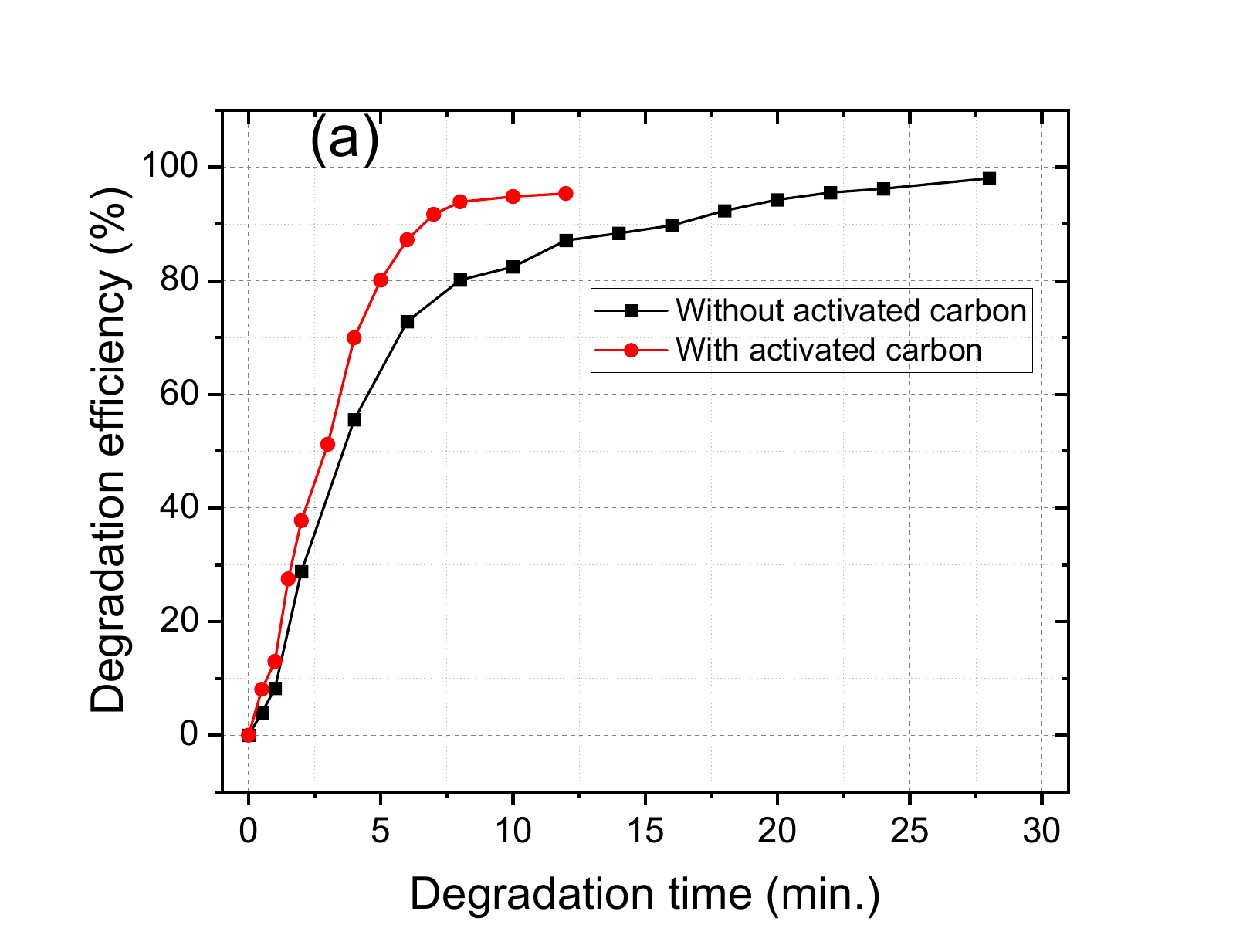}
    \end{minipage}%
    \hfill
    \begin{minipage}[t]{0.33\textwidth}
        \centering
        \includegraphics[width=\linewidth]{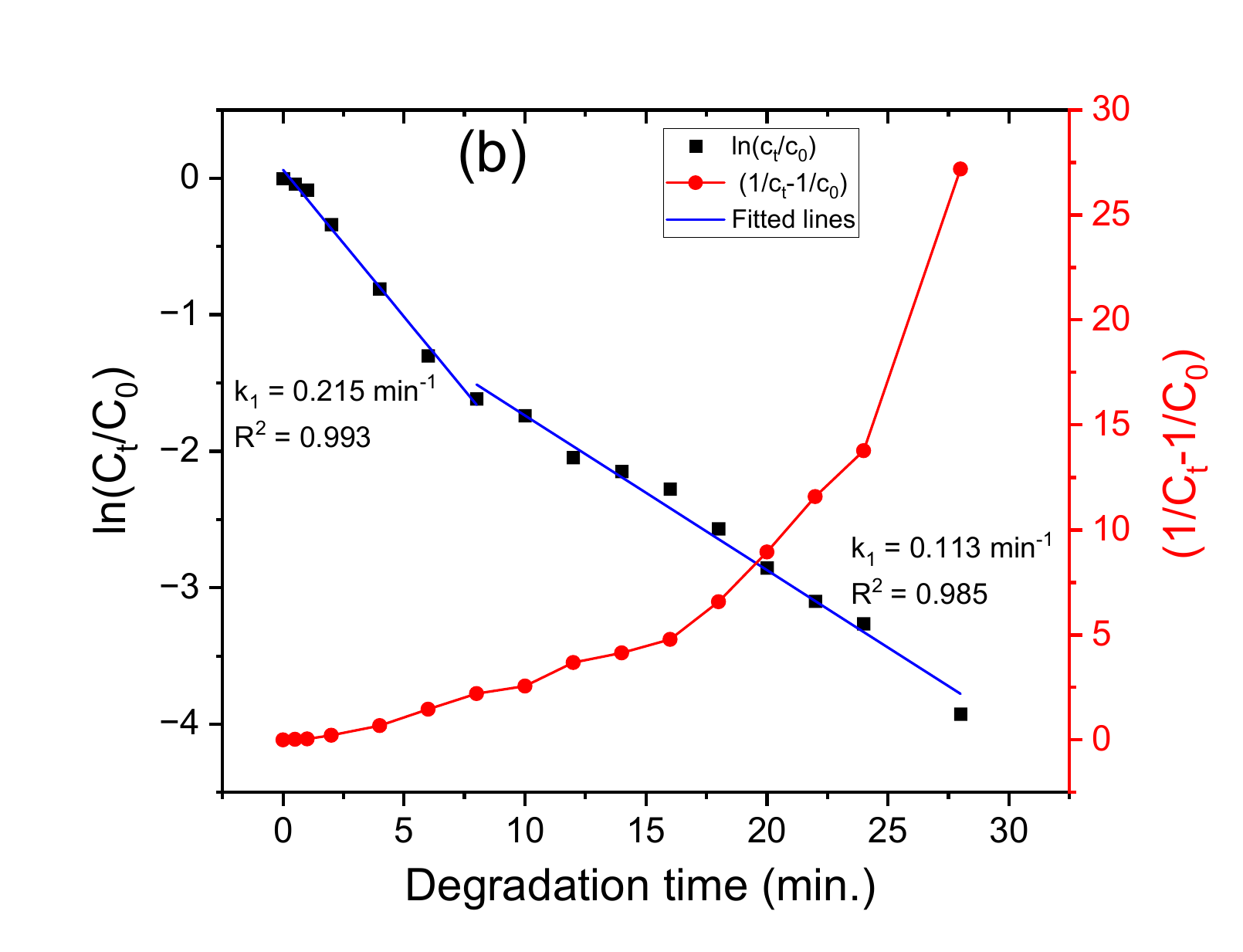}
    \end{minipage}%
    \hfill
    \begin{minipage}[t]{0.33\textwidth}
        \centering
        \includegraphics[width=\linewidth]{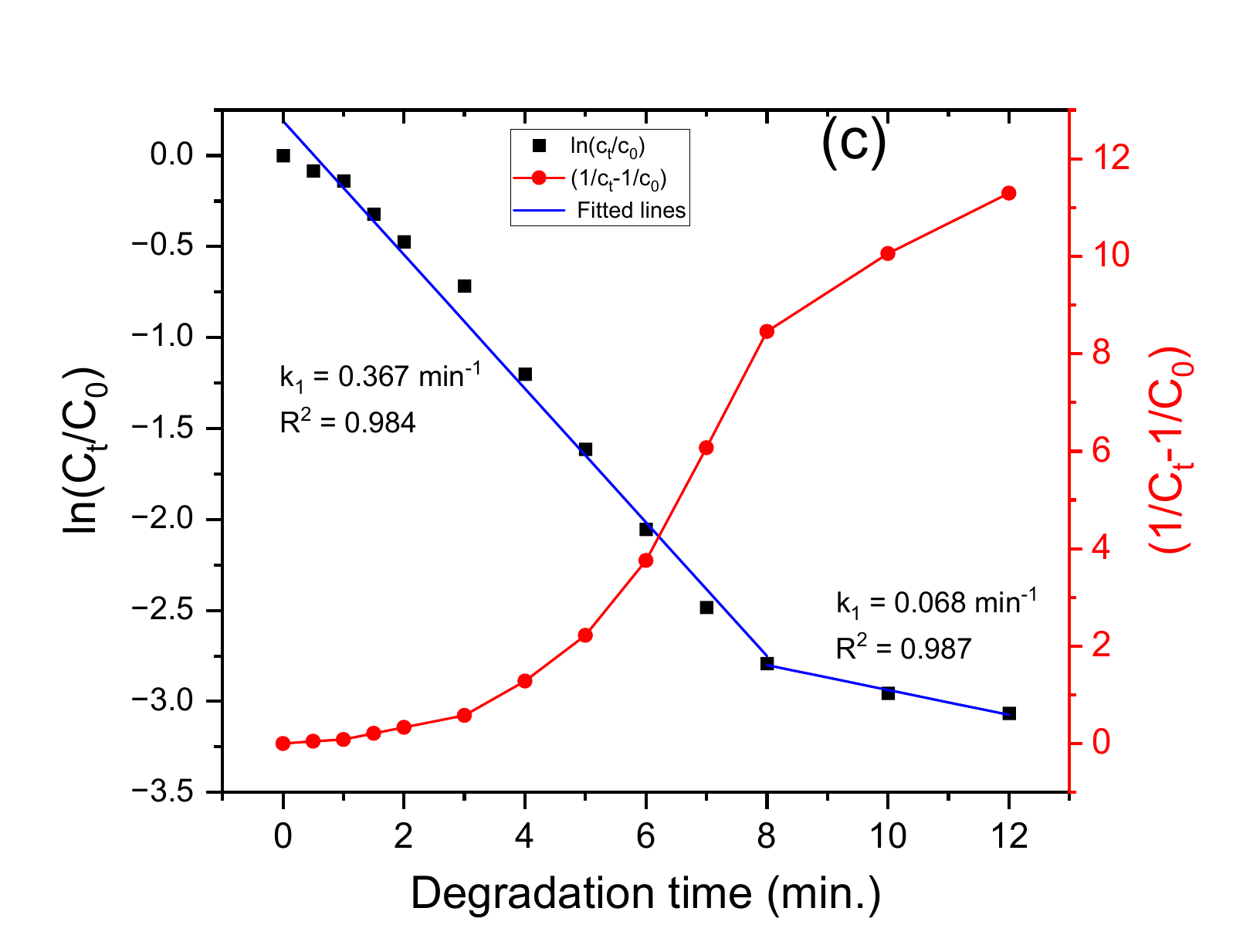}
    \end{minipage}

    \caption{{(a) (a) Degradation efficiency of the symmetric ternary anionic dye mixture (30 mg/L) under alkaline conditions with and without AC. The degradation reaction kinetics of this solution, (b) in the absence of activated carbon, and (c) in the presence of activated carbon. The solution volume, gas flow rate, and magnetic stirrer speed were 800 ml, 300 cc/min, and 400 RPM, respectively.}}
    \label{fig:Fig7}
\end{figure*}

To extend our studies, further experiments were conducted to degrade an 800 ml symmetric ternary anionic dye mixture (30 mg/L) under basic conditions in the main experimental chamber, using similar experimental conditions. For attaining a basic pH ($\sim$ 11.7), a diluted solution of KOH (0.1 M) was used. To study the catalytic effect of activated carbon on basic dye solutions, a ternary anionic dye mixture was degraded in both the presence and absence of AC. Unlike the acidic conditions (Fig.~\ref{fig:Fig6}(a)), the dye mixture (30 mg/L) with alkaline (basic) nature took more time to degrade (Fig.\ref{fig:Fig7}(a)) both in the presence ($\sim$ 12 min.) and absence of AC ($\sim$ 28 min.). It is clear from Fig.\ref{fig:Fig7}(a) that more than 95 \% of dyes degrade in nearly 10 minutes of treatment with AC, while about 80 \% of dyes decompose in the same time in the absence of AC. To achieve more than 95 \% degradation of dyes in a ternary mixture, around 30 minutes post-discharge treatment is required in the absence of AC. It demonstrates a significant role of AC as a catalyst in decolorization (degradation) of anionic dye mixtures having an alkaline nature.\\\\
From the UV-Vis absorption spectrum data, the reaction kinetics of the basic ternary dye system (with and without AC) were analyzed. The concentration functions against treatment time for first and second order degradation reactions for both cases (with and without AC) are plotted in Fig.\ref{fig:Fig7}(b) and Fig.\ref{fig:Fig7}(c). In the absence of AC,  the fitted plots of $\ln\left(\frac{C_t}{C_0}\right)$ against the degradation time in Fig.\ref{fig:Fig7}(b) indicate the two-phase first-order reaction kinetics rather than the second-order degradation reaction. The fragmentation of the dye molecules occurs in two phases (0-8 and 8-30 minutes). During the first phase of fragmentation, the reaction kinetics of degradation take place with a larger rate constant (k$_{1}$ = 0.215 $min^{-1}$) compared to that (k$_{1}$ = 0.113 $min^{-1}$) of the second degradation phase. In the presence of AC, as shown in Fig.\ref{fig:Fig7}(c), the dye degradation follows first-order reaction kinetics instead of zero or second order, with two different phases. The decomposition of the anionic dye molecules occurs in two phases (0-8 and 8-12 minutes). During the first phase of fragmentation, the degradation takes place with a larger rate constant (k$_{1}$ = 0.367 $min^{-1}$) compared to that of the second degradation phase (k$_{1}$ = 0.068 $min^{-1}$). The dye degradation in basic solution (with and without AC) with two different first-order phases could be due to various factors, such as the formation of intermediate organic compounds, the absorption of degraded fragments onto the AC surfaces, and different reaction pathways in the presence of smaller-sized organic compounds. \\\\
In line with the objectives of investigating the decolourization of an anionic dye mixture under different environments (acidic, neutral, or basic), a series of experiments was performed with and without activated carbon. The degradation time ($t$) during which more than 90 \% of dye molecules decompose into smaller fragments during post-discharge treatment with and without activated carbon is presented in Table~\ref{table1}. We observe a slight difference in degradation time between solutions with and without activated carbon when the treated anionic dye solution is either acidic or neutral. The decolourization time of the same mixture solution increases with increasing alkaline strength (higher pH). However, adding AC as a catalyst during treatment increases the decomposition rate by nearly threefold, thereby reducing the degradation time by a similar margin.
\begin{table*} 
\caption{A comparative list of anionic dye degradation (between 90 to 95 \%) time and energy yield in the absence and presence of activated carbon. Gas flow rate was 300 cc/min, magnetic stirrer speed was 400 RPM, and volume of the anionic dye solution was 800 ml.}
\label{table1}
  \centering  
\begin{tabular}{|p{2cm}|p{3cm}|p{3cm}|p{3cm}|p{3cm}|p{3cm}|} \hline
 \bfseries pH of dye solution & \bfseries Concentration of dye solution & \multicolumn{2}{|c|}{\bfseries Degradation time (minutes)}&\multicolumn{2}{|c|}{\bfseries Energy yield of dye degradation}\\ \hline
 & \bfseries  (mg/L) & \bfseries  Without AC & \bfseries  With AC & \bfseries Without AC (g/kWh) & \bfseries With AC (g/kWh)\\ \hline
1.49 & 15  & 3.0 $\pm$ 0.5 & 3.5  $\pm$ 0.33  & 7.862 $\pm$ 0.41 & 7.147 $\pm$ 0.38\\ \hline
6.95 &  15   & 4 $\pm$ 0.66 & 4.2 $\pm$ 0.33 & 5.922 $\pm$ 0.31 & 6.08 $\pm$ 0.32\\ \hline
 11.7 & 15 & 15 $\pm$ 1 & 8 $\pm$ 0.5 & 1.707 $\pm$ 0.09 & 3.211 $\pm$ 0.17\\ \hline
 9.6 & 30 & 8 $\pm$ 0.5 & 8.5 $\pm$ 0.66 & 6.423 $\pm$ 0.34 & 5.96 $\pm$ 0.31\\ \hline
 11.8 & 30 & 32 $\pm$ 3 & 12 $\pm$ 2 & 1.564 $\pm$ 0.08 & 3.914 $\pm$ 0.2\\ \hline
 12.30 & 30 & 52 $\pm$ 3 & 22 $\pm$ 2  & 0.995 $\pm$ 0.05 & 2.28 $\pm$ 0.13\\ \hline
\end{tabular}
\end{table*}
\subsection{Role of concentration of anionic dye mixture}
\begin{figure}[t]
    \centering
    \begin{minipage}[t]{0.47\textwidth}
        \centering
        \includegraphics[width=\linewidth]{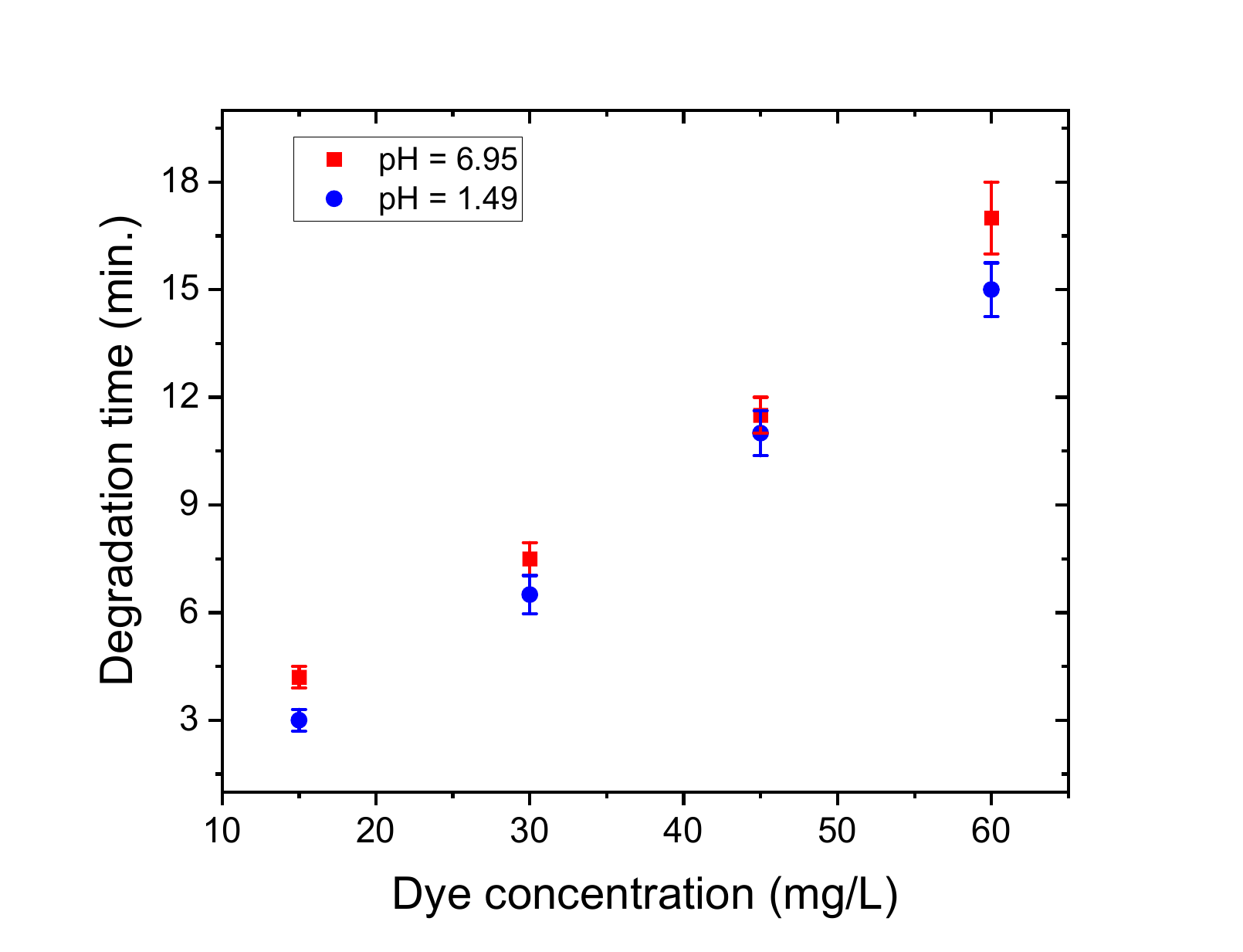}
    \end{minipage}%
    \hfill
 \caption{Variation of degradation time versus concentration of symmetric anionic dye mixture at gas flow rate of 300 cc/min and magnetic stirrer speed of 400 RPM.}
    \label{fig:Fig8}
\end{figure}

The initial dye concentration is a critical parameter that influences the efficiency of the DDBD plasma reactor by affecting the dye degradation rate. To obtain the relation between concentration of anionic ternary dye mixture (symmetric) and degradation ($>$ 95 \%) time, experiments were conducted in neutral (pH = 6.95) and strong acidic (pH $=$ 1.49) solutions of different concentrations under similar experimental conditions (flow rate, 300 cc/min and stirrer speed, 400 RPM) without activated carbon. An approximate linear relationship between the concentration of dye solutions (synthetic dye wastewater) and degradation time was observed across different pH values (see Fig.\ref{fig:Fig8}). The degradation efficiency is concentration-dependent (in any environment), with a higher initial dye load requiring longer treatment times to achieve equivalent decomposition efficiencies. We also observe similar trends in degradation time versus concentration in the presence of AC, under basic or acidic conditions in synthetic dye wastewater.  
\begin{table*} 
\centering
\caption{Effect of cumulative AC loading on plasma treatment time. pH of solution, concentration of dye mixture, volume of solution, gas flow rate and magnetic stirrer speed were 11.8, 30 mg/L, 800 ml, 300 cc/min, 400 RPM, respectively.}
\label{Table2}
\begin{tabular}{|c|c|}
\hline
\textbf{Activated carbon loading } & \textbf{Post-discharge treatment time (min)} \\
\hline
2g (fresh) & 20 $\pm$ 2 \\
2g (fresh) + 2 g ($1^{st}$ used) & 16 $\pm$ 1.5\\
2g (fresh) + 4 g ($1^{st}$ and $2^{nd}$ used) & 14 $\pm$ 1\\
2g (fresh) + 6 g ($1^{st}$, $2^{nd}$ and $3^{rd}$ used) & 13 $\pm$ 1\\
\hline
\end{tabular}
\end{table*}
\begin{table*}   
\centering
\caption{Effect of AC reuse on plasma treatment time. Amount of AC was 2g, pH of solution was 11.8, concentration of dye mixture was 30 mg/L, volume of solution was 800 mL, and glow flow rate was 300 cc/min.}
\label{Table3}
\begin{tabular}{|c|c|}
\hline
\textbf{Activated carbon usage cycle} & \textbf{Post-discharge treatment time (min)} \\
\hline
2g fresh AC & 21 $\pm$ 2 \\
2g $1^{st}$ reuse & 12 $\pm$ 1 \\
2g $2^{nd}$ reuse & 10.5 $\pm$ 1 \\
2g $3^{rd}$ reuse & 10 $\pm$ 1 \\
2g $4^{th}$ reuse & 11 $\pm$ 1 \\
\hline
\end{tabular}
\end{table*}
\subsection{Effect of activated carbon loading and usage cycle}
The effect of activated carbon (granules) mass loading (amount of AC) was investigated by cumulatively adding fresh and reused activated carbon under alkaline conditions (pH 11.8) to the anionic ternary dye mixture (30 mg/L, 800 ml). By increasing the AC loading from 2 grams to 8 grams by adding 2g in each degradation experiment, the decolourization time reduces from $\sim$ 20 to 13 minutes (Table~\ref{Table2}). The effect of AC loading becomes insignificant after a certain amount (mass) of activated carbon as a catalyst in enhancing the dye degradation rate (see Table~\ref{Table2}). In the present case, we observe little change in the degradation rate once the activated carbon load exceeds 4 grams.  However, these experiments validate the catalytic role of activated carbon in enhancing anionic dye degradation in the alkaline environment.\\ 
The reusability of activated carbon was investigated through multiple treatment cycles (experiments) of ternary anionic dye solutions (30 mg/L, pH = 11.8). The degradation times of dye mixtures using 2 grams of AC across different experiments are presented in Table~\ref{Table3}. It was observed that reused AC performed better than fresh AC, with the shortest treatment time obtained during the third reuse cycle. The dye decomposition time was $\sim$ 20 minutes in the case of fresh AC, while the decomposition time was reduced by nearly 50 \% in the third cycle of dye solution treatment with the used AC (4g), which was used in the first and second cycles of the experiments (Table~\ref{Table3}). Therefore, the reused AC remained more effective than fresh AC, demonstrating good regeneration potential and practical applicability for repeated plasma treatment cycles. 
\subsection{pH and EC of solution during treatment}
The pH and electrical conductivity of the samples collected for UV–Vis measurements under acidic and basic conditions, both in the presence and absence of activated carbon (AC), were monitored as given in Table~ \ref{tab:table4}.  The acidic conditions (pH 1.30--1.42) exhibited excellent stability throughout the treatment process, with a standard deviation of  0.04 units for both the A-NAC (acidic pH without AC) and A-AC (acidic pH with AC) groups. The conductivity remained relatively constant in the ranges of 7.01--7.93 mS/cm for A-NAC and 7.95--8.05 mS/cm for A-AC, indicating only minor changes in the solution's ionic composition during treatment. In the case of basic anionic ternary dye solutions, the pH remained nearly constant during the treatment.  Conductivity remained relatively constant in the range 860--935 $\mu$S/cm, and, as shown in Table~\ref{tab:table4}, the plasma treatment produced only minor changes in the ionic composition of the solution. 
\begin{table*}
\centering
    \caption{Variation of pH and electric conductivity (EC) for anionic dye solutions with different solution conditions.}     
    \label{tab:table4}
    \begin{small}
    \begin{tabular}{|c|c|c|c|c||c|c|c|c|}
    \hline
    Time & \multicolumn{2} {|c|} {\bfseries Acidic sol. (without AC)} & \multicolumn{2} {|c|} {\bfseries Acidic (with AC)} & \multicolumn{2} {|c|} {\bfseries Basic sol. (without AC)} & \multicolumn{2} {|c|} {\bfseries Basic (with AC)}\\
    \cline{2-9}
    
    {\bfseries minutes } & {\bfseries pH} &{\bfseries EC ($mS/cm$)}  & {\bfseries pH} &{\bfseries EC ($mS/cm$)} & {\bfseries pH} &{\bfseries EC ($\micro S/cm)$} & {\bfseries pH} & {\bfseries EC ($\micro S/cm)$)}  \\
    \hline
    0       & 1.35 $\pm$ 0.20  & 7.01 $\pm$ 0.30  & 1.39 $\pm$ 0.17  & 7.95 $\pm$ 0.30 &  11.83 $\pm$ 0.15  & 923 $\pm$ 12 & 11.79 $\pm$ 0.20 & 902 $\pm$ 10 \\
    \hline
    2        & 1.32$ \pm$ 0.20  & 7.83 $\pm$ 0.25 & 1.38 $\pm$ & 8.05 $\pm$ 0.30&  11.79 $\pm$ 0.20  &  916 $\pm$ 10 & 11.77 $\pm$ 0.18 & 906 $\pm$ 13\\
    \hline
    5          & 1.31 $\pm$ 0.22    & 7.93 $\pm$ 0.33  & 1.39 $\pm$ 0.2  & 8.01 $\pm$ 0.30 &  11.81 $\pm$ 0.20 & 915 $\pm$ 10 & 11.76 $\pm$ 0.20 & 918 $\pm$ 10\\
    \hline
    8      & 1.39 $\pm$ 0.16 & 7.89 $\pm$ 0.27 & 1.42 $\pm$ 0.15 & $7.95 \pm$ 0.250&  11.83 $\pm$ 0.18 &  901 $\pm$ 15 & 11.75 $\pm$ 0.15 & 935 $\pm$ 15 \\
    \hline
    14        & -& - & -  & - &   11.79 $\pm$ 0.20 & 860 $\pm$ 15 & 11.72 $\pm$ 0.16 & 930 $\pm$ 15\\
    \hline
    20         & -   & - & - & -  &   11.78 $\pm$ 0.17 & 895 $\pm$ 16 & - &-\\
    \hline
    28      & -  &  - & - & -&  11.75 $\pm$ 0.15 & 903 $\pm$ 15 &- &-\\
    \hline
    \end{tabular}
    \end{small} 
\end{table*}
\subsection{Comparative degradation at different treatment stages}
\begin{figure}[t]
    \centering

    \begin{minipage}[t]{0.47\textwidth}
        \centering
        \includegraphics[width=\linewidth]{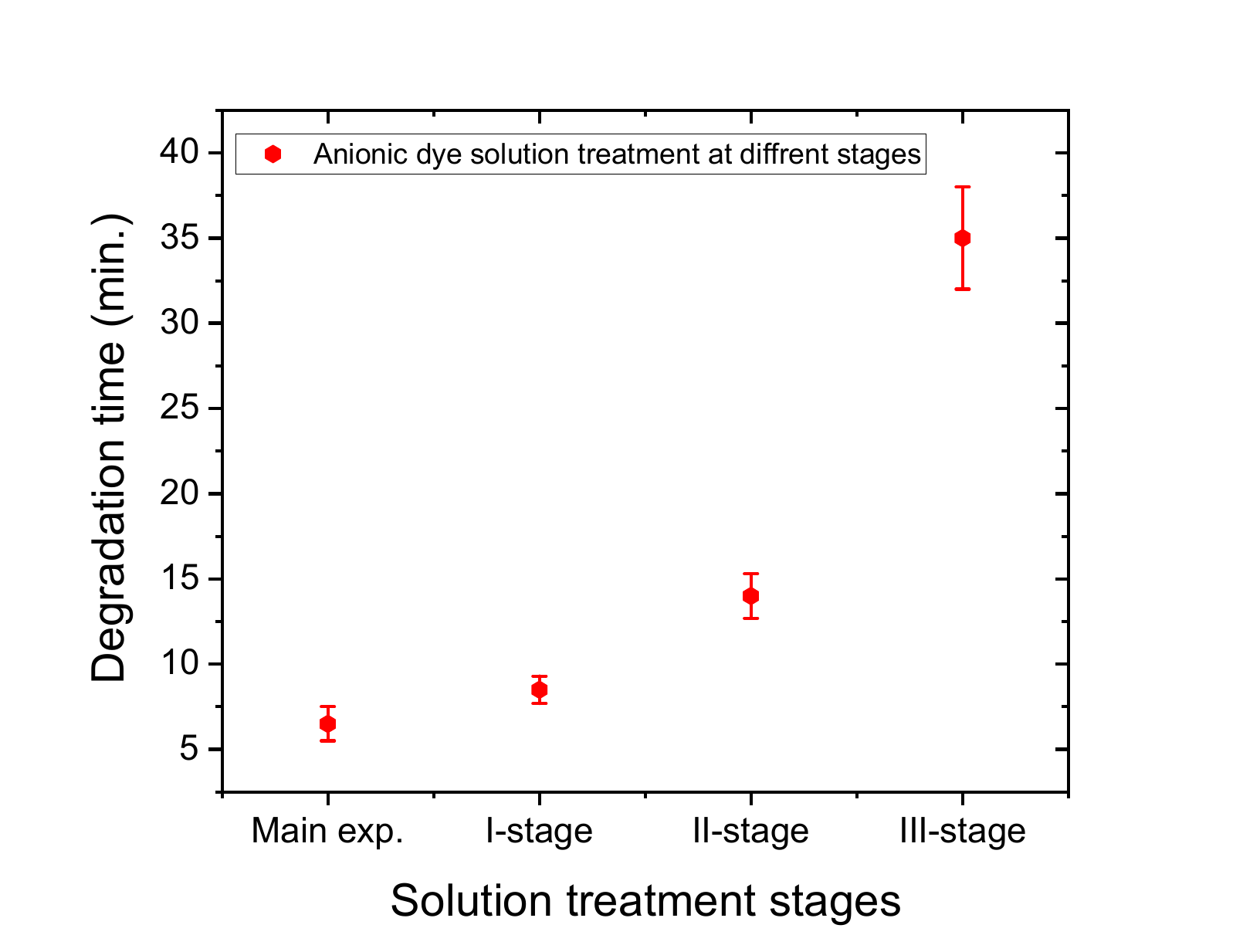}
    \end{minipage}%
    \hfill
    \caption{Degradation of the symmetric ternary anionic dye mixture (30 mg/L) at different stages (main experimental chamber, stages I, II and III). The gas flow rate was 300 cc/min, and the magnetic stirrer speed was 400 RPM.}
    \label{fig:Fig9} 
\end{figure}

In the present study, we conducted experiments in the main experimental vessel (stage I) and allowed the undissolved emanating reactive gases, along with oxygen, to pass through the sequential stages (stage II, stage III, and stage IV), as per the schematic diagram (see Fig.\ref{fig:Fig1}).  
The use of a multistage process for synthetic wastewater treatment, rather than a single stage, was intended to maximize the utilization of reactive species for simultaneous degradation of different solutions and to reduce the exposure of ozone (a major reactive species) to the environment. To explore the degradation energy yield of the dual-liquid post-DDBD reactor, a set of experiments was performed using a symmetric ternary anionic mixture (pH $\sim$ 7), with equal concentrations (30 mg/L) and volumes (800 ml) in each container at different treatment stages. The dyes decolourization time at different stages is plotted in Fig.\ref{fig:Fig9}.  We observe a nearly increasing exponential trend of decolourization time. This is due to the reduction in oxidizer mass transfer rate with increasing numbers of treatment stages, thereby increasing the dye degradation time from stage I to stage IV. The energy yield was calculated for an average dissipated power of 25 W and a degradation efficiency of 95 \%. The energy yield decreases from the main experimental vessel (8.42 g/kWh) followed by stage I (6.44 g/kWh), II (4.05 g/kWh) and III (1.48 g/kWh). So, the net energy yield with multistage treatment increases from 8.42 g/kWh (single stage) to 20.39 g/kWh. [Another study was conducted to elaborate on the benefits of using multi-stages in such post-discharge treatment].  
\section{Discussion} \label{sec:sec5}
The results obtained in this study clearly demonstrate the effectiveness of the dual-liquid post-dielectric barrier discharge plasma reactor in degrading anionic dyes, which are considered synthetic dye wastewater, under both acidic and basic conditions. Since dry oxygen with 99 \% purity (industrial grade) was used as the input gas (feed gas) to the reactor for plasma formation, ozone ($O_3$) was assumed to be the major long-lived reactive oxygen species, along with small traces of nitrogen species. Therefore, we measured the concentration of ozone in post-discharge reactive gases using iodometric titration\cite{mangi_post-discharge_2026}. \begin{figure}[t]
    \centering
    \begin{minipage}[t]{0.47\textwidth}
        \centering
        \includegraphics[width=\linewidth]{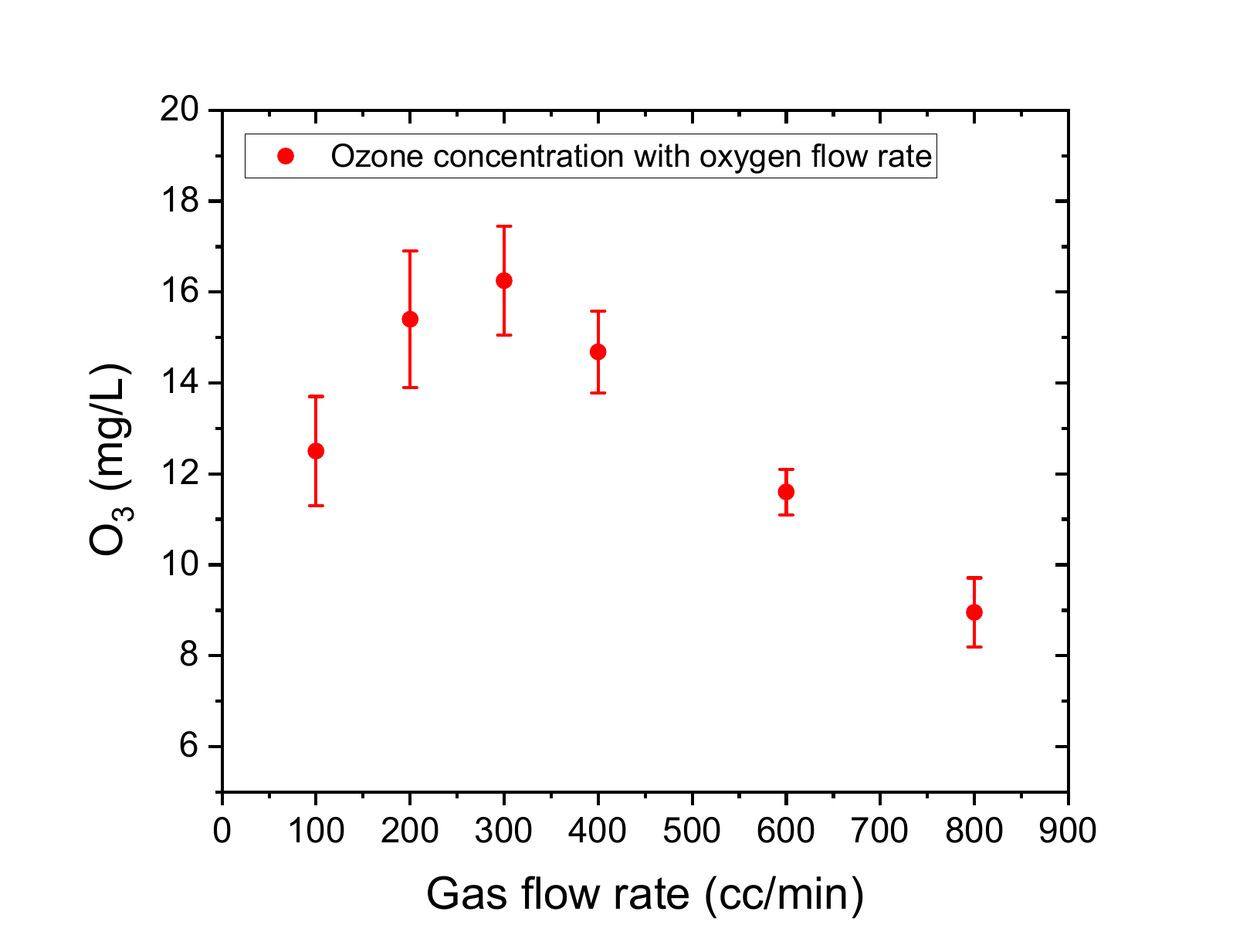}
    \end{minipage}%
    \hfill
    \caption{Ozone concentration in reactive emanating gases from the dual-liquid DDBD reactor at different oxygen flow rates.}
    \label{fig:Fig10}
\end{figure}
The concentration of $O_3$ measured at different oxygen flow rates is plotted in Fig.\ref{fig:Fig10}. 
The $O_3$ concentration increases from $\sim$ 12.5 mg/L at 100 cc/min to a maximum value of $\sim$ 16 mg/L at a gas flow rate between 200 cc/min and 400 cc/min, reflecting a more efficient conversion of oxygen to ozone during the discharge period. Further increases in gas flow rates ($>$ 400 cc/min) show a decreasing trend in ozone concentration, reaching approximately 9.0 mg/L at 800 cc/min. This reduction in $O_3$ concentration is expected due to the shorter residence time of oxygen molecules at higher gas flow rates, which limits oxygen dissociation and subsequent ozone formation. These measurements guided us in performing dye degradation experiments (in the present work) using the newly developed dual-liquid DDBD plasma reactor in the post-discharge configuration, at flow rates between 300 cc/min and 400 cc/min, to achieve a higher degradation energy yield.\\
There is a possibility for the presence of hydrogen peroxide (H$_2$O$_2$) as a long-lived oxidizer in the emanating reactive gases from the DDBD reactor if water molecules are present in the discharge zone. In the present study, we used dry oxygen in the reactor, so the formation of H$_2$O$_2$ in the discharge zone is unexpected. However, there is a possibility of forming H$_2$O$_2$ after passing reactive gases to dye solutions. In aqueous medium, ozone ($O_3$) converts into hydroxyl radicals ($\cdot OH$ ) through a chain reaction. Then $H_2O_2$ is usually formed through the recombination of hydroxyl radicals\cite{ohradical} as per the reaction
\begin{equation}
\mathrm{\cdot OH + \cdot OH \rightarrow H_2O_2} 
\end{equation}
The presence of H$_2$O$_2$ is used as an indirect indicator of hydroxyl radical generation during the post-discharge treatment of wastewater solutions. To measure the concentration of H$_2$O$_2$ (indirectly $\cdot OH$) in the anionic dye solutions during post-discharge treatment, a chemical test kit was used, and the maximum concentration was measured at 5 mg/L throughout the treatment period. This confirms that no significant accumulation of hydrogen peroxide occurred in the liquid phase during exposure to post-discharge gases. It also confirms that the degradation mechanism is dominated by ozone rather than $\cdot OH$.\\ 
 Nitrite (NO$_2^-$) and nitrate (NO$_3^-$) ions are commonly produced during the breakdown of nitrogen and oxygen gases present in the discharge zone. The reaction pathway\cite{rns} can be represented as:
\begin{equation}
\mathrm{N_2 + O_2 \rightarrow NO \rightarrow NO_2 \rightarrow HNO_2/HNO_3}
\end{equation}
Measurements of the concentration of these reactive nitrogen species in the reactive emanating gases using test kits revealed no detectable nitrite (NO$_2^-$) and the presence of nitrate (NO$_3^-$) at 5 ppm after 30 minutes of treatment. The lack of NO$_2^-$ together with the presence of NO$_3^-$ suggests that nitrogen oxidation proceeded to the stable nitrate form due to its higher oxidation state and fully delocalized electron structure. The small trace of nitrogen species in the dye solution is due to the presence of small amounts of $N_2$, along with $O_2$ in the input feed gas. Therefore, the role of nitrogen reactive species in the dye degradation process is assumed to be negligible.  
\begin{figure*}[t]
    \centering
    \includegraphics[width=0.95\textwidth]{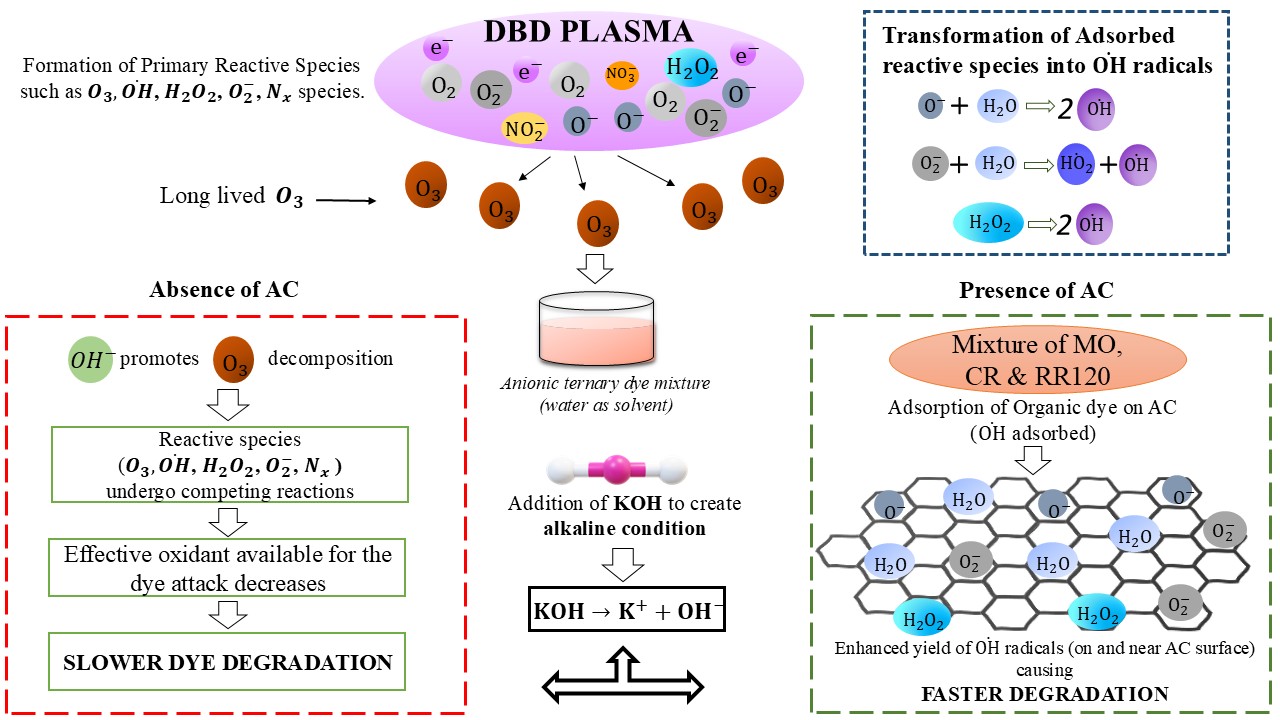}
    \caption{{Flowchart representing the role of activated carbon (catalyst) in degrading anionic dye molecules in alkaline solution.}}
    \label{fig:Fig11}
\end{figure*}
\\
In the emanating reactive gases (post-discharge), $O_2$ is always present along with $O_3$ and nitrogen reactive species. Since $O_2$ itself is an oxidizer, to differentiate the effects of oxygen alone from the plasma-generated reactive species ($O_3$), an experiment was performed in which pure oxygen was continuously dissolved into a ternary dye mixture (30 mg/L) without operating the DDBD reactor. A longer treatment does not produce any noticeable change in the dye solution's colour. This confirms the negligible role of $O_2$ relative to ozone in the decolourization reaction pathways during such treatment in the absence of a suitable catalyst.\\
Thus, in the present study, ozone is a major oxidizer responsible for the degradation (decolourization) of the anionic dyes. It either directly attacks dye molecules (organic matter) or generates free radicals (via indirect reactions). The proportion ratio of each reaction mechanism (direct or indirect) depends on various factors \cite{ozonation_azodye_reactions,ozonation_dye-degradation_path2,ozonereaction_pathways, Ozonation_2024}. Under acidic conditions in an anionic dye mixture, ozone acts directly as an electrophile on organic substrates (dyes). The probability of formation of free radicals is negligible; therefore, degradation of anionic dyes is mainly initiated by dissolved $O_3$ through direct reaction in the presence and absence of activated carbon as a catalyst. In addition to $O_3$, the dissolved oxygen in acidic ternary anionic solutions can also form hydronium ($H_{3}O^{+}$) ions, which act as an oxidizing agent. The possible reaction pathways\cite{dissociation_sulfuric_acid,dissociation_sulfuric_acid2,dissociation_sulfuric_acid3} are given as:- 

\begin{align}
\mathrm{H_{2}SO_{4} + H_{2}O \;\longrightarrow\; H_{3}O^{+} + HSO_{4}^{-}} \label{eq:first-step} \\
\mathrm{HSO_{4}^{-} + H_{2}O \;\rightleftharpoons\; H_{3}O^{+} + SO_{4}^{2-}} \label{eq:second-step}
\end{align}
where the equation(\ref{eq:first-step}) and (\ref{eq:second-step}) represent complete and partial dissociation respectively.
The hydronium ($H_{3}O^{+}$) ions formed here serves as an oxidizing agent\cite{h3o_oxidiser} and is expected to convert the dissolved oxygen ($O_{2}$) into ozone ($O_{3}$) in the solution. This led to the promotion of direct oxidation by ozone and other oxidizers, resulting in an attack on the azo bond. Therefore, we observe slightly higher degradation efficiency for the acidic (pH $<$ 4) ternary anionic dye mixture (synthetic dye wastewater), as shown in Table~\ref {table1}. In the presence of activated carbon, we observe a slight increase (insignificant) in the degradation time compared to that without activated carbon (see Table~\ref {table1}). This can be due to the complexity of the dye system or due to some chemical reactions \cite{AC_sulfuric, AC_sulfuric2} of AC with $H_{3}O^{+}$ ions or any other components formed by the combination of MO, CR and RR120 with sulfuric acid. \\
In a basic dye solution (without AC), the dissociated $OH^{-}$ promotes ozone (O$_{3}$) decomposition, and O$_{3}$ is relatively unstable (compared to acidic conditions), leading to the competing reaction of reactive species ($\dot{O}H$, $\dot{O}_{2}^{-}$, $H\dot{O}_{2}$). Due to these competing reactions of oxidizers and their scavengers (reaction inhibitors), the quantity of effective oxidants available for the fragmentation of long dye molecules decreases, which reduces the degradation rate. As we know, anionic dye molecules carrying negative charges due to sulfonate groups ($SO_3^{-}$) may experience an electrostatic repulsive force in the basic solutions, which could also slow the degradation rate\cite{Malarvizhi2010TheIO,DESOUZA2018662}. \\
To enhance dye degradation under alkaline conditions, several experiments \cite {h2o2_treatment} (addition of $H_{2}O_{2}$, iron powder, etc.) and studies were conducted. When AC granules (2g) were added to the dye mixture in the main experimental chamber (with a magnetic stirrer speed of 400 RPM), faster dye degradation was recorded (see Table~\ref{table1}) for higher pH value solutions. This reduction in treatment time is expected due to the adsorption of ozone and reactive species onto the active sites of highly porous granular AC, as well as the transformation of adsorbed reactive species into $\dot{O}H$ radicals. A typical degradation reaction pathway in the presence of AC in the alkaline anionic dye solution is presented in Fig.\ref{fig:Fig11}. During treatment, dye molecules approach or adsorb onto the AC surface, thereby enhancing interfacial oxidation. The azo bonds and aromatic structures are attacked by adsorbed oxidizers, resulting in faster fragmentation of dye molecules. \\
To check whether activated carbon particles adsorb dye molecules during post-discharge treatment, a control experiment was conducted to evaluate the contribution of adsorption alone to the AC surfaces in an alkaline dye solution. In the absence of plasma treatment, we did not observe any noticeable change in the dye solution colour after stirring it with a 6 g activated carbon load. It clearly indicates that the adsorption of dye molecules was insufficient for significant dye removal under the investigated conditions. 
\section{Summary and future work} \label{sec:sec6}
The reported work in this article is summarized as:
\begin{itemize}
    \item The dual-liquid double dielectric barrier discharge (DDBD) plasma reactor for plasma-based synthetic wastewater treatment has been successfully built and operated in a post-discharge configuration.
    \item The plasma-generated reactive long-lived oxidizers (mainly $O_3$) in the post-discharge configuration are responsible for the oxidation and breakdown of long chains of dye molecules through direct or indirect reaction paths into smaller molecules.
    \item Coconut-shell-derived granular activated carbon acted as a catalyst and significantly enhanced the degradation efficiency of anionic ternary dye mixture (alkaline environment) by adsorbing primary reactive species ($O_{3}$ \& $H_{2}O_{2}$) on its highly porous surface. It promoted the formation of hydroxyl radicals ($\dot{OH}$) under alkaline conditions, which is the strongest oxidizer. 
    \item In acidic or neutral anionic ternary mixture solutions, the role of granular activated carbon in determining the degradation efficiency is insignificant, indicating that degradation is dominated by ozone through direct reaction.
    \item The degradation process is influenced by operating parameters such as pH of solution, activated carbon (AC) mass loading (2-6 g), re-usability of AC, and its particle sizes and shapes.
    \item The degradation reaction kinetics exhibited zero, first, and second order with two phases (slow and fast degradation) depending on the solution environment (acidic or basic) and the presence or absence of activated carbon (catalyst) during the post-discharge treatment.
\end{itemize}

Although the present study demonstrated the effectiveness of the low-cost dual-liquid post-DDBD plasma reactor for dye degradation (simulated dye wastewater), further investigations are required to evaluate the reactor's performance in treating realistic textile industrial wastewater containing various dyes, salts, and other organic contaminants. The degradation study (at the lab scale) of realistic industrial textile wastewater, along with potential catalysts and oxidation processes, would help identify challenges to the adaptability of plasma technology in wastewater treatment. It is also necessary to use additional analytical techniques (which can be made possible by the expansion of laboratory facilities) to estimate COD, BOD, TOC, and toxicity to better understand the extent of pollutant mineralization. 
\section{Acknowledgement} This work is supported by the Science and Engineering Research Board (SERB) [Note: SERB has recently been subsumed by the Anusandhan National Research Foundation (ANRF)], a statutory body of the Department of Science \& Technology (DST), Government of India, under a Start-up Research Grant (SRG) with the sanction number: SRG/2023/000757. The authors sincerely thank the Department of Physics and Astrophysics for allowing us to use the Central Experimental Facilities (CEF) for UV-Vis spectroscopy. The authors are also grateful to the physics workshop staff at the Department of Physics and Astrophysics, University of Delhi, for their assistance in developing this dual-liquid post-DDBD reactor.     
\section{Author Declarations}
\subsection{Conflict of Interest}
The authors declare no conflicts of interest related to this article.
\subsection{Author Contributions}
Dr Mangilal Choudhary conceived the idea for this project and, with Krishna's assistance, developed the dual-liquid DDBD plasma reactor. The experimental work on the degradation of anionic dye mixture under all possible conditions was performed by Miss Krishna. Mr Surya assisted in experiments and spectroscopic analysis of dye solutions. The first draft of the manuscript was written by Mangilal and Krishna, and all authors revised it. All authors read and approved the final manuscript.   
\section{Data Availability}
The datasets generated during and/or analyzed in the current study are available from the corresponding author upon reasonable request.
\bibliography{biblography}

\end{document}